\documentclass[a4paper,fleqn]{cas-sc}

\usepackage{enumerate}
\usepackage[authoryear,longnamesfirst]{natbib}
\usepackage{graphicx,amssymb,amsmath,layout,verbatim,rotating,calc}
\newcommand{\ej}{E$_{J}$ }
\def\tsc#1{\csdef{#1}{\textsc{\lowercase{#1}}\xspace}}
\tsc{WGM}
\tsc{QE}
\tsc{EP}
\tsc{PMS}
\tsc{BEC}
\tsc{DE}

\begin{document}
\let\WriteBookmarks\relax
\def\floatpagepagefraction{1}
\def\textpagefraction{.001}
\shorttitle{ $\Delta - Chaos$ relation}
\shortauthors{P.A.~Patsis et~al.}


\title [mode = title]{Chaoticity in the vicinity of complex unstable periodic
orbits in galactic type potentials}




\author[1]{P.A.~Patsis}[]
\cormark[1]
\fnmark[1]
\ead{patsis@academyofathens.gr}
\ead[url]{astro.academyofathens.gr/people/patsis/}


\address[1]{Research Center for Astronomy, Academy of Athens, Soranou
Efessiou 4, 115 27 Athens, Greece}

\author[2]{T.~Manos}[]

\ead{thanos.manos@cyu.fr}
\ead[URL]{https://sites.google.com/site/thanosmanos}
%

\address[2]{Laboratoire de Physique Th{\'e}orique et Mod{\'e}lisation, CY Cergy
Paris Universit{\'e}, CNRS, UMR 8089, 95302 Cergy-Pontoise cedex, France}

\author[3,4,5]{L.~Chaves-Velasquez}
\ead{leonardochaves83@gmail.com}

\address[3]{Astronomical Observatory, University of Nari\~{n}o, Sede VIIS,
Avenida Panamericana, Pasto, Nari\~{n}o, Colombia}
\address[4]{Departamento de F\'{\i}sica de la Universidad de Nari\~{n}o,
Torobajo Calle 18 Carrera 50, Pasto, Nari\~{n}o, Colombia}
\address[5]{Instituto de Radioastronom\'{\i}a y Astrof\'{\i}sica, Universidad
Nacional Aut\'{o}noma de M\'{e}xico, Apdo.Postal 3-72, Morelia, Michoac\'{a}n,
58089, Mexico}

\author[6]{Ch.~Skokos}[orcid=0000-0002-2276-8251]
\ead{haris.skokos@uct.ac.za   }
\ead[url]{http://math_research.uct.ac.za/~hskokos/}


\address[6]{Nonlinear Dynamics and Chaos group, Department of Mathematics and
Applied Mathematics, University of Cape Town, Rondebosch, 7701, South Africa}

\author[7]{I.~Puerari}[]
\ead{puerari@Inaoep.mx}
\ead[url]{https://www.inaoep.mx/~puerari/}


\address[7]{Instituto Nacional de Astrof\'{\i}sica, Optica y Electr\'onica,
Calle
Luis Enrique Erro 1, 72840, Santa Mar\'{\i}a Tonantzintla, Puebla, Mexico}

\cortext[cor1]{Corresponding author}


\begin{abstract}
We investigate the evolution of phase space close to complex unstable periodic
orbits in two galactic type potentials. They represent characteristic
morphological types of disc galaxies, namely barred and normal (non-barred)
spiral galaxies. These potentials are known for providing building blocks to
support observed features such as the peanut, or X-shaped bulge, in the former
case and the spiral arms in the latter. We investigate the possibility that
these structures are reinforced, apart by regular orbits, also by orbits in the
vicinity of complex unstable periodic orbits. We examine the evolution of the
phase space structure in the immediate neighbourhood of the periodic orbits in
cases where the stability of a family presents a successive transition from
stability to complex instability and then to stability again, as energy
increases. We find that we have a gradual reshaping of invariant structures
close to the transition points and we trace this evolution in both models.
We conclude that for time scales significant for the dynamics of
galaxies, there are weakly chaotic orbits associated with complex unstable
periodic orbits, which  should be considered as structure-supporting, since they
reinforce the morphological features we study.

%
\end{abstract}




\maketitle

\section{Introduction}


Complex instability is a particular type of orbital instability that
appears in autonomous Hamiltonian systems of three or more degrees of
freedom (for a definition see Sect.~\ref{defi}). In galactic dynamics it
characterizes periodic orbits of many
three dimensional (hereafter 3D) models in a large volume of their
parameter space \citep{m82a, m82b, pf84, pf85b, gcom85, gco86, mpf87, pf87,
mdz88, z88, pz90, z93, pz94, opf98, kpc11, pk14a}. However,
considerable insight in the role of complex instability for the
dynamics of Hamiltonian systems has been gained by works on several
other kinds of potentials \citep{h85, pcp95, op99} or 4-dimensional
symplectic mappings \citep{pf85a, cogi88, opf95, jo04, deco16,
sb21}. These results have also been used for understanding the
behaviour of stellar orbits in galactic type potentials.

A main problem in galactic dynamics is to find the orbital building
blocks that can reinforce observed morphological features in
galaxies. In that respect, finding stable periodic orbits, which
attract around them quasi-periodic orbits that remain in their
neighbourhood, is essential for understanding the enhancement of bars
and spiral arms in 3D autonomous, rotating, Hamiltonian systems representing
disc galaxies. Nevertheless, the quasi-periodic orbits must provide the
appropriate shapes that match the morphology of the structure we study.

A particular structure can also be supported during a certain time by
orbits that remain confined during this period in a specific volume of the
phase space. Such orbits are called sticky and have been mainly studied in
two-dimensional systems \citep[see][and references therein]{ch10}. Much less
work has been done on stickiness in 3D models and especially in the
neighbourhood of complex unstable periodic orbits. The present study
investigates whether we can find orbits that remain confined close to periodic
orbits characterized by this type of instability.

Complex instability has been considered as an abrupt transition to
chaos, since its appearance is not associated with the introduction of new
families of periodic orbits in the system. In any case, studies of the phase
space in the neighbourhood of complex unstable periodic orbits have shown that
the transition from stability to complex instability gives rise to bifurcating
invariant structures in Poincar\'{e} sections of 3D autonomous Hamiltonian
systems  \citep{pf85b, pcp95, opf98, oetal04, kpc11}, as well as in 4D
symplectic maps \citep{pf85a, jo04, zkp13}.

A question that arises, is if the degree of chaoticity of orbits close
to a complex unstable periodic orbit can be associated with some quantity that
can indicate the presence or absence of invariant forms in the corresponding
phase space region. From the analysis leading to the stability of the periodic
orbits, the obvious candidate is the discriminant $\Delta$ (for the definition
see Sect.~\ref{defi} below). One of the goals of the present paper is to
investigate its relation with chaoticity in phase space, as quantified by chaos
indicators like the GALI$_2$ index, introduced by \citet{sba07}. However, the
ultimate goal of the study is to explore whether or not one can find close to
complex unstable periodic orbits building blocks for supporting structures like
the two main morphological features of disc galaxies, which are the bars and the
spiral arms.

The paper is structured as follows: In Section~\ref{defi} we present the
various kinds of instabilities of periodic orbits in 3D systems. In
Section~\ref{models} we describe the two dynamical models, which we have used in
our study, namely the Ferrers bar and the PERLAS potential. In
Section~\ref{gali} we give definitions associated with the GALI$_2$ indicator.
Our numerical results  are presented in Section~\ref{results}.  Finally we
enumerate our conclusions in Section~\ref{concl}.

\section{Orbital instabilities in 3D systems}
\label{defi}

In galactic disc dynamics we deal with disc-like potentials, rotating around an
axis perpendicular to the disc, at the center of the system. In such a case,
our
Hamiltonian can be described in Cartesian coordinates $(x,y,z)$ as:

\begin{equation}
\label{ham}
H= \frac{1}{2}(p_{x}^{2} + p_{y}^{2} + p_{z}^{2}) +
    V(x,y,z) - \Omega_{s}(x p_{y} - y p_{x})    ,
\end{equation}
where $V(x,y,z)$ is the potential of the model, $\Omega_{s}$ the rotational
velocity of the system (pattern speed), which is constant and $p_{x},~ p_{y},$
and $p_{z}$ are the canonically conjugate momenta. The axis of rotation is the
$z$ axis.

We will refer to the conserved numerical value of the Hamiltonian,
$E_J$, as the Jacobi constant or, more loosely, as the `energy'.

The equations of motion corresponding to Eq.~\ref{ham} are:
\begin{eqnarray}
\dot{x}=p_{x}+\Omega_{s}y, & \dot{y}=p_{y}-\Omega_{s}x, &
\dot{z}=p_{z} \nonumber \\
\dot{p_{x}}= -\frac{\partial V}{\partial x} + \Omega_{s}p_{y}, &
\displaystyle{\dot{p_{y}}=-\frac{\partial V}{\partial y} - \Omega_{s}p_{x},} &
\dot{p_{z}}=-\frac{\partial V}{\partial z}.
\end{eqnarray}

The space of section in the case of a 3D system is 4D. The equations of motion
for a given $E_J$ are solved numerically, starting with initial conditions
$(x_{0},z_{0},p_{x_0},p_{z_0})$ in the plane $y$=0 (with the $p_{y_0}$ value
determined by the given \ej value) and then by  considering
successive upwards ($p_y > 0$) intersection with this plane.

The exact initial conditions for the periodic orbit are calculated using a
Newton iterative method. A periodic orbit is found when the initial and final
coordinates coincide with an accuracy of at least 10$^{-10}$. The integration
scheme used was a 4th order Runge-Kutta algorithm and in some cases a
Runge-Kutta Fehlberg 7-8th order scheme, securing a relative error in the
energy less than $10^{-14}$.

Resonances play a crucial role in the dynamics of rotating, 3D, galactic
potentials. These are resonances between the epicyclic and the rotational
frequencies of the stellar orbits, in the rotating with $\Omega_s$ frame of
reference (radial resonances), while we have vertical frequencies as well, in
which, instead of the epicyclic, we consider the vertical frequency. A special
case is the corotation resonance, where the angular velocity of the stars is
equal to the pattern speed, namely $\Omega(r)$ = $\Omega_s$ \citep[for
definitions see e.g.][]{pg96}. For the needs of the present study we keep in
mind that at the resonances the stability of a family of periodic orbits, i.e.
of periodic solutions of the equations of motion (2), may change \citep{cg89}.

When a periodic orbit is found, it can be characterized as stable or unstable by
calculating its linear stability. This is done by following a  method introduced
by \citet{b69} and \citet{h75}. \citet{gcom85} have distinguished three
kinds of instability for the unstable periodic orbits. The method is briefly
described below, where we also give the definitions of the instabilities.

The first step is to consider a small deviation from the initial conditions of
the periodic orbit and then to integrate the perturbed orbit again up to the
next upward intersection. In this way a 4D Poincar\'{e} map, $T: \mathbf{R}^{4}
\to \mathbf{R}^{4}$, is established, relating the points of initial with the
final deviation. In vector form this relation can be written as:
$\vec{\xi}=M\,\vec{\xi_{0}}$, where $\vec{\xi}$ is the final deviation,
$\vec{\xi_{0}}$ is the initial deviation and $M$ a $4 \times 4$
matrix, called the monodromy matrix.
The characteristic equation of this matrix is written in the form $\lambda^{4}
+
\alpha
\lambda^{3} + \beta \lambda^{2} + \alpha \lambda + 1 = 0$.
Its solutions $(\lambda_i, i=1,2,3,4)$, obey the relations
$\lambda_{1}\,\lambda_{2}=1$ and $\lambda_{3}\,\lambda_{4}=1$ and we can write
for each pair:
\begin{equation}
\lambda_{i}, 1/\lambda_{i} = \frac{1}{2}
\left[-b_{i}\pm(b_{i}^{2} -4)^{\frac{1}{2}}\right],
\end{equation}
where $\displaystyle b_{i} = \frac{1}{2}\,( \alpha \pm \Delta^{1/2})$ and
\begin{equation}
\label{delta}
\Delta =\alpha^{2} - 4 (\beta - 2).
\end{equation}
Stability or Instability of the periodic orbit is
expressed by means of the quantities $b_{1}, b_{2}$ and $\Delta$.
The quantities $b_{1}$ and $b_{2}$ are called the
stability indices. One of them is associated with radial and the other one with
vertical perturbations. We distinguish the following four cases:
\begin{enumerate}[(1)]
 \item If $\Delta > 0$, $|b_{1}|<2$ and $|b_{2}|<2$, the 4
eigenvalues $\lambda_{i} (i=1,2,3,4)$ are on the unit circle and the periodic
orbit is called
`stable', (S).
\item If $\Delta > 0$, and $|b_{1}|>2$, $|b_{2}|<2$, or
$|b_{2}|>2$, $|b_{1}|<2$, two eigenvalues are on the real axis and two
on the unit circle, and the periodic orbit is called `simple
unstable', (U).
\item If $\Delta > 0$, $|b_{1}|>2$, and $|b_{2}|>2$, all four
eigenvalues are on the real axis, and the periodic orbit is called
`double unstable', (DU).
\item Finally, $\Delta < 0$ means that all four
eigenvalues are complex numbers but {\em off} the unit circle. The
orbit is characterized then as ``complex unstable'', ($\Delta$).
\end{enumerate}

For a general discussion of the kinds of instability encountered in
Hamiltonian systems of {\sf N} degrees of freedom the reader may refer
to \citet{s01}.

As one of the parameters of our model varies (in this work $E_J$), case (4) may
appear at an S$\rightarrow\!\! \Delta$ or at a DU$\rightarrow\!\! \Delta$
transition. When the periodic orbit is initially stable, we have at a critical
$E_J$, a pairwise collision of eigenvalues on two conjugate points of the unit
circle. Then, Krein-Moser theorem \citep[see e.g.][p.298]{gcobook} decides if
the eigenvalues will remain on the unit circle after the collision, or if they
will move out of the unit circle,  into the complex plane, forming a complex
quadruplet. In the former case the orbits of the family will continue being
stable and in the latter they will become complex unstable. The transition from
stability to complex instability is also known as Hamiltonian Hopf Bifurcation
\citep{vdM85}.


At complex instability, unlike in the two other kinds of instabilities,
we do not have  introduction of new  families of periodic orbits in the system.
In the case of a S$\rightarrow$U transition, the stability of the parent family
is inherited to a bifurcated one. Thus, for values of the parameter beyond the
critical value for which the stability of the family changes, new tori of
quasiperiodic orbits will appear in the phase space of the system, belonging to
the new stable families. The U$\rightarrow$DU transition occurs to a later stage
and can be considered as a transition from order to chaos in two steps
(S$\rightarrow$U$\rightarrow$DU). A bifurcated family at the U$\rightarrow$DU
transition will be simple unstable. At a DU$\rightarrow\!\!\Delta$ transition no
new families are introduced in the system. We note that in the latter case the
neighbourhood of the parent family is already chaotic and no new invariant
structures are encountered in phase space \citep{kpc11}.

\section{The Dynamical Models}
\label{models}

Complex instability appears frequently, as energy varies, in the evolution of
the stability of the main 3D families of periodic orbits that bifurcate from the
central, planar family of periodic orbits x1, and make up the ``x1-tree''
\citep{spa02}. The most important of these families is x1v1, which bifurcates
from x1, usually as stable \citep[but see also][]{ph18}, at the
vertical 2:1 resonance. The existence of the x1-tree families is not associated
with a particular model, but it is an intrinsic property of any 3D rotating
potential, in which the resonances can be defined. They offer the building
blocks for supporting the main structures encountered in disc galaxies, namely
the bar \citep[see e.g.][]{psa02} and the spirals \citep{pg96, chvv19}.

\subsection{Ferrers bar}
\label{ferrers}
The first potential we used, refers to a mass distribution representing a
galactic bar. The 3D bar is rotating around its short $z$ axis. The $x$ axis is
the intermediate and the $y$ axis the long one. The system is rotating with the
pattern speed of the bar $\Omega_{b}$, i.e. $\Omega_s = \Omega_b$. The bar is
embedded in a 3D disc, while in the center of the system exists also a
spheroidal bulge. Thus, this galactic model consists of three components, a
disc, a
bulge and a bar.

The disc is represented by a Miyamoto potential \citep{mn75}:
\begin{equation}
\label{potd}
\Phi _{D}=-\frac{GM_{D}}{\sqrt{x^{2}+y^{2}+(A+\sqrt{B^{2}+z^{2}})^{2}}},
\end{equation}
where \( M_{D} \) is the total mass of the disc, $A$ and $B$ are the
horizontal and vertical scale lengths, and $G$ is the gravitational
constant.

The bulge is modelled by a Plummer sphere with potential:
\begin{equation}
\label{pots}
\Phi _{S}=-\frac{GM_{S}}{\sqrt{x^{2}+y^{2}+z^{2}+\epsilon _{s}^{2}}},
\end{equation}
where \( \epsilon _{s} \) is the scale length of the bulge and \(
M_{S} \) is its total mass.

The third component of the potential is a
triaxial Ferrers bar, whose density \( \rho \) is:
\begin{equation}
\label{densd}
\rho =\left\{ \begin{array}{lcc}
\displaystyle{\frac{105M_{B}}{32\pi abc}(1-m^{2})^{2}} & {\mbox for} &
m \leq 1\\
 & & \\
\displaystyle{0} & {\mbox for}  & m>1
\end{array}\right. ,
\end{equation}
where
\begin{equation}
\label{semiaxis}
m^{2}=\frac{y^{2}}{a^{2}}+\frac{x^{2}}{b^{2}}+\frac{z^{2}}{c^{2}}\, \, ,\, \,
 \,
 a>b>c,
\end{equation}
\( a \), \( b \), \( c \) are the semi-axes and \( M_{B} \) is the mass of the
bar component. The corresponding potential \( \Phi _{B} \) and the forces are
given in a closed form in \citet{pf84}\footnote{We made use of the offer of
\citet{opf98} for free access to the electronic version of the potential and
forces routines.}.  We use for the Miyamoto disc A=3 and B=1, and for the
Ferrers bar axes we set $a\!:\!b\!:\!c$ = $6\!:\!1.5\!:\!0.6$, as in
\citet{pf84} and in many previous works of the authors using this potential. The
masses of the three components satisfy \( G(M_{D}+M_{S}+M_{B})=1 \). The length
unit is taken as 1~kpc, the time unit as 1~Myr and the mass unit as $ 2\times
10^{11} M_{\odot}$. In Table~\ref{tab:models} we give the parameters of our
model. They have been chosen so, that for this model we have a typical
alternation of stable and complex unstable regions in the x1v1 family, as the
energy varies.
\begin{table*}
\caption[]{The parameters of the Ferrers bar model: G is the
  gravitational constant, M$_D$, M$_B$, M$_S$ are the masses of the
  disc, the bar and the bulge respectively, $\epsilon_s$ is the scale
  length of the bulge, $\Omega_{b}$ is the pattern speed of the bar,
  $E_J$(v-IILR) is the value of the Jacobi constant for
  the vertical 2:1 resonance and $R_c$ is the corotation
  radius.}
\label{tab:models}
\begin{center}
\begin{tabular}{ccccccccc}
  GM$_D$ & GM$_B$ & GM$_S$ & $\epsilon_s$ & $\Omega_{b}$ &
$E_J$(v-ILR) & $R_c$ \\
\hline
  0.87 &  0.05  & 0.08 & 0.4 &  0.054 & -0.3028 & 6.38 \\

\hline
\end{tabular}
\end{center}
\end{table*}

\subsection{PERLAS spirals}
\label{perlas}
The second potential we used, refers to a bisymmetric logarithmic spiral as
those observed in grand design spiral galaxies. Spirals are, besides the bars,
the second feature that characterizes the morphology of disc galaxies.
They
may appear together with a bar (barred-spiral galaxies) or without a bar
(normal spiral galaxies). The
spiral potential is embedded in an axisymmetric background that has three parts.
The first two parts are represented by the same general models we used for the
axisymmetric components of the bar model. Namely, we have first a central mass
component, $\Phi_{S}$, representing the bulge, given again by Eq.~\ref{pots},
with mass $M_{S}$ and scale length $\epsilon_{s}$. In addition, we consider a
3D Miyamoto disc (Eq.~\ref{potd}) with mass $M_{D}$ and scale lengths $A$, $B$.

The third component of the axisymmetric part of the spiral potential refers to
a massive halo, represented by a halo potential proposed by \citet{as91},
which at radius $r$ is given by
\begin{eqnarray}
\Phi_{H}(r)=-\left(\frac{M(r)}{r}\right)-\left(\frac{M_{H}}{1.02a_{H}}
\right)\times\nonumber
 \left[-\frac{1.02}{1+(r/a_{H})^{1.02}}+\ln(1+(r/a_{H})^{1.02})\right]^{100}_{r}
,
\end{eqnarray}
where
\begin{equation}
 M(r) = \frac{M_{H}(r/a_{H})^{2.02}}{1+(r/a_{H})^{1.02}}.
\end{equation}
$M(r)$ has mass units, $M_{H}$ is the mass of the halo, and $a_{H}$ is a
scale length.

The perturbation in this case has the form of a three dimensional spiral
component, for which we use the PERLAS (sPiral arms potEntial foRmed by obLAte
Spheroids) potential \citep{bp03}. The spiral
pattern has two arms and is shaped by a density distribution formed by
individual, inhomogeneous, oblate Schmidt spheroids \citep{sch56}, These
spheroids are superposed along a logarithmic spiral locus of constant pitch
angle $i$. The spirals are considered to be trailing and rotating clockwise.
%

									

The Schmidt spheroids have constant semi-axes ratio, while their density falls
linearly outwards, starting from their centres on the spiral locus. The
separation among the centres of the spheroids is 0.5 kpc, their total width
2~kpc and their total height 1~kpc, The spiral arms start at 2.03~kpc and end at
12.9~kpc. These distances correspond to the radial 2:1 resonance and to 1.5
times the corotation radius respectively. The density along the loci of the
spiral arms falls exponentially, as the one of the disc does. For a detailed
presentation of the PERLAS potential see e.g. \citet{pvetal12}.


The total potential is the sum of the four terms: $\Phi = \Phi_{S} + \Phi_D +
\Phi_H + \Phi_{SP}$, the three first of which compose the axisymmetric part and
$\Phi_{SP}$ is the PERLAS spiral potential. The parameters for all these
components used in the particular model of the present paper are summarized in
Table~\ref{tab:sparameters}.

\begin{table*}
\begin{center}
\centering \footnotesize \caption{Parameters of the PERLAS potential. The upper
row refers to the spiral part, while the two lower rows give the values of all
parameters of the axisymmetric components (see text).}
\label{tab:sparameters}
\scalebox{0.8}{
\begin{tabular}{ccccccc}

\multicolumn{7}{c}{Spiral part} \\
\cline{1-2}

\hline\\

Galaxy type    & Locus & Arms Number & Pitch Angle $i^{o}$ & $\mu=M_{SP}/M_{D}$
& Scale length (kpc) & $\Omega_{p}$ ($\text{km}/\text{s}/\text{kpc}$)\\[0.2cm]
  Sc    & Logarithmic    & $2$  & $25^{o}$ & $0.04$ & $3.7$ & $-20$    \\

\hline\\

\multicolumn{7}{c}{Axisymmetric Components} \\
\cline{1-2}

\hline\\

$M_{D}/M_{H}$    & $M_{S}/M_{D}$ & Maximum of rotation velocity
($\text{km}\text{s}^{-1}$) & $M_{D}(10^{10}M_{\odot})$ &
$M_{S}(10^{10}M_\odot)$ & $M_{H}(10^{11}M_\odot)$ & Disc Scale length (kpc)
\\[0.2cm]
$0.1$     & $0.2$    & $170$  & $5.10$ & $1.02$ & $4.85$ & $3.7$      \\
\hline\\

 & $\epsilon_{s}$ (kpc) & $A$ (kpc) & $B$ (kpc) & $a_H$ (kpc) &  \\ [0.2cm]
 &   1         &   5.32    &  0.25     &    12       &  \\
\hline
\end{tabular}
}
\end{center}
\end{table*}
The pattern speed $\Omega_p=-20$~km s$^{-1}$ kpc$^{-1}$ defines the angular
velocity of the system in the clockwise sense, while the pitch angle of the
logarithmic spirals, $i=25^{\circ}$, corresponds to an open pattern, typical of
galaxies of galactic type Sc. The model has a maximum rotational
velocity of 170~$\text{km}\text{s}^{-1}$, that is typical for a galaxy of
morphological type Sc \citep{retal80}. The amplitude of the  perturbation is
determined
by the ratio $M_{SP}/M_{D}$, where $M_{SP}$ is the mass of the spiral and
$M_{D}$ the mass of the disc component \citep{chvv19}. This set up has been
chosen for the needs of the present paper, since the stability of the basic 3D
family x1v1, which supports the spiral arms \citep{chvv19}, has successive
stable and complex unstable parts, as \ej varies.
We note that the scaling of units in the two models is not the same, so the
numerical values referring to the Ferrers bar and the PERLAS model are
different.

\section{The GALI$_2$ indicator}
\label{gali}
In order to quantify the chaoticity of the orbits we consider in the present
study, we use the standard chaos indicator GALI$_2$ \citep{sba07}.

The GALI$_2$ index is given by the norm of the wedge product of two
normalized to unity deviation vectors $\mathbf{\hat{w}}_{1}(t)$ and
$\mathbf{\hat{w}}_{2}(t)$: $\rm{GALI}_{2}(t)=|\mathbf{\hat{w}}_{1}(t)\wedge
\mathbf{\hat{w}}_{2}(t)|.$ The initial coordinates of the deviations vectors are
chosen randomly, and the two vectors are orthonormalized by using the
Gram-Schmidt process at the beginning of the integration. This sets the initial
value of the index to $\mbox{GALI}_2(0)=1$. Thus, in order to evaluate GALI$_2$
we integrate the equations of motion and the variational equations for two
deviation vectors simultaneously. The GALI$_2$ index behaves as follows
\citep[see][and references therein]{sm16}:

\begin{itemize}
\item For
chaotic orbits it falls exponentially to zero as:
$\mbox{GALI}_{2}(t)\propto \exp{(-(\lambda_{1}-\lambda_{2})t)}$,
where $\lambda_{1}$ and $\lambda_{2}$ are the two largest Lyapunov exponents
\citep[for definitions and for the computation of the Lyapunov exponents see
e.g:][]{ben80,s10}.
\item For regular orbits it
oscillates around a positive value across the integration:
$\mbox{GALI}_{2}(t)\propto constant$.

\item In the case of
sticky orbits we observe a transition from practically
constant GALI$_2$ values,
which correspond to the seemingly quasiperiodic
epoch of the orbit, to an
exponential decay to zero, which indicates the
orbit's transition to
chaoticity.
\end{itemize}

GALI$_2$ has been used to characterize the chaoticity of the orbits, both in
Ferrers bars \citep{ma11} as well as in PERLAS potentials \citep{chvv19}.

\section{Results}
\label{results}
\subsection{Complex Unstable regions in Ferrers bars}
\label{deltafer}
\begin{figure}
\centering
\includegraphics[scale=0.3]{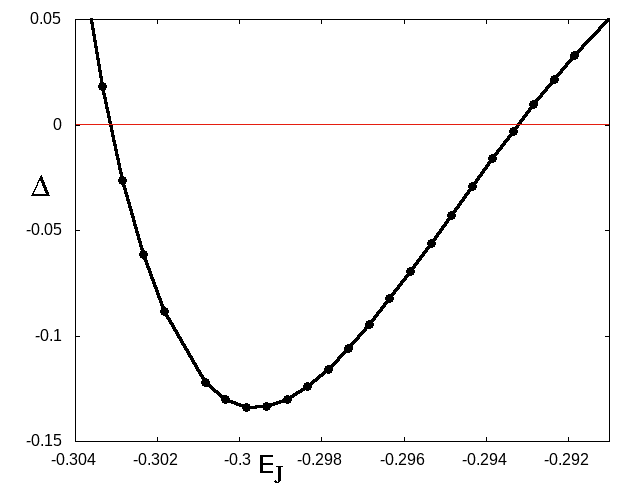}
\caption{Ferrers bar: The variation of $\Delta$ (Eq.~\ref{delta}) for the x1v1
family of periodic
orbits in the
$-0.304 \leq E_J \leq -0.291$ region}. The heavy dots correspond to calculated
periodic
orbits. Those with $\Delta < 0$ are complex unstable.
\label{ejdel_005a}
\end{figure}

We have studied the chaoticity in the neighbourhood of complex unstable periodic
orbits belonging to the family x1v1 \citep{spa02} in two energy regions of our
Ferrers bars model. The orbits of this family are important, because they act as
building blocks for the peanut-shaped bulges in the central regions of barred
galaxies \citep{psa02}. In the first case we studied, the complex unstable
region is found between a S$\rightarrow\!\!\Delta$ transition at \ej$\approx
-0.3028$ and a $\Delta\!\! \rightarrow$S transition at \ej$\approx -0.293$. At
the critical \ej values, there is a sign change of $\Delta$, being $\Delta < 0$
in the complex unstable region. In Fig.~\ref{ejdel_005a} we give the
variation of $\Delta$ in the $-0.304 \leq E_J \leq -0.291$ interval.

\begin{figure}
\centering
\includegraphics[scale=0.87]{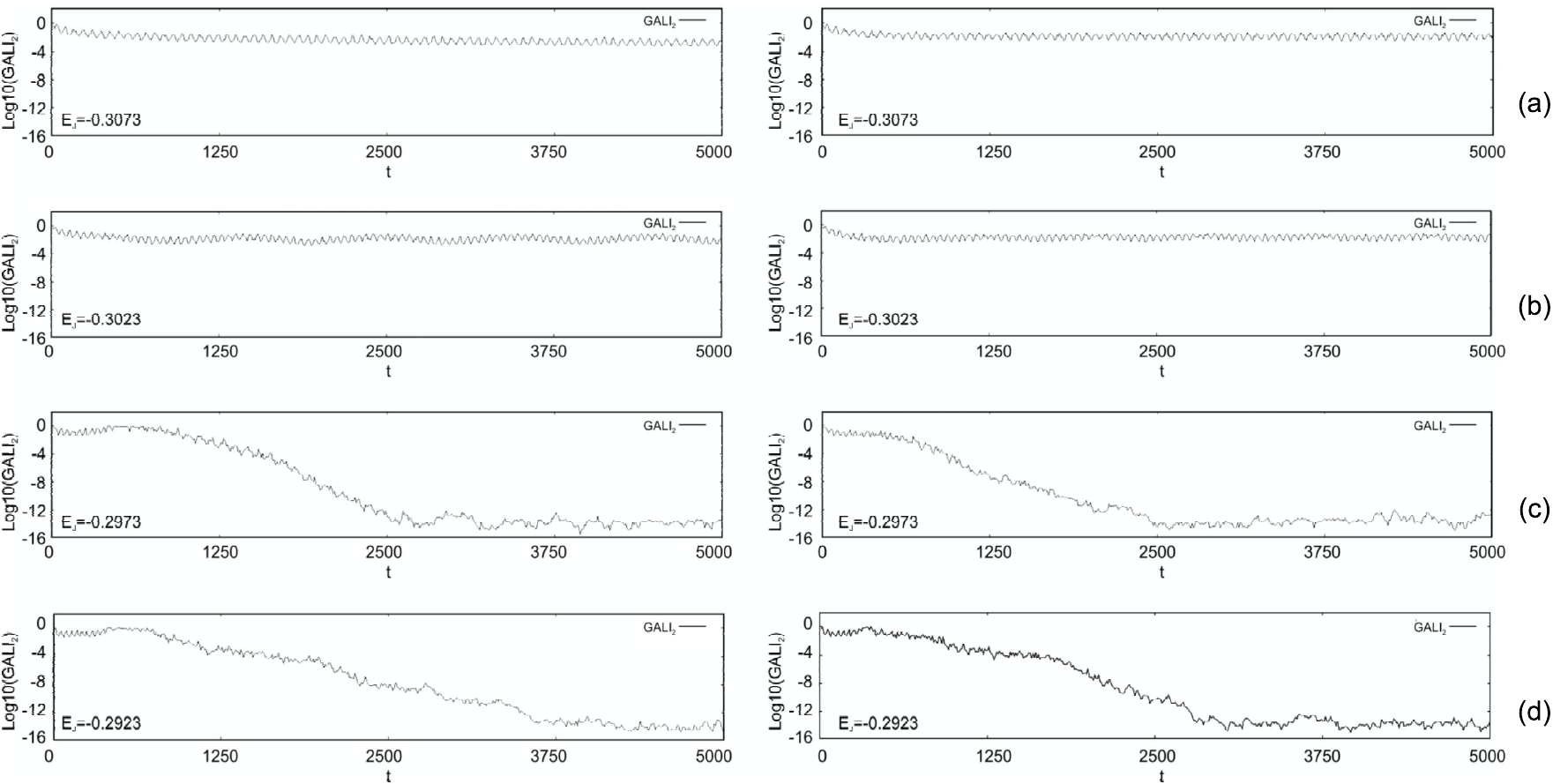}
\caption{Ferrers bar: The GALI$_2$ indices of orbits in the first
S$\rightarrow\Delta\rightarrow$S transition, for
which we have calculated $\Delta$ in Fig.~\ref{ejdel_005a}. In the left column
the orbits have x1v1 initial conditions perturbed in the x-direction by 10\% of
$x_0$, while in the right column the orbits are again x1v1, but perturbed this
time in the z-direction by 10\% of $z_0$ (see text). In (a) and (d) the
corresponding periodic orbit is stable (S), while in (b) and (c) complex
unstable ($\Delta$).
In (d), the perturbed orbit close to a stable x1v1 has a
chaotic behaviour. The integration time corresponds to 5~Gyr.}
\label{galisd1}
\end{figure}

The quantity $\Delta$ (Eq.~\ref{delta}) refers to the periodic orbit itself. In
order to find out
whether, and how, it is associated with the phase space structure in the
neighbourhood of x1v1, we perturbed the initial conditions of the
orbits of this family at different \ej\!. We first investigated orbits with
initial conditions those of the periodic orbit, with one of the coordinates
perturbed by 10\% of its value. We consider this as a reasonable perturbation
of a periodic orbit for finding non-periodic orbits that could potentially
participate in the reinforcement of the peanut-shaped bulge. In particular, at
each \ej\!\!, we calculated first the GALI$_2$ index of an orbit with the
initial conditions of x1v1, perturbed in the $x$-direction by $0.1
x_0(\rm{x1v1})$ and then the GALI$_2$ index of an orbit with the initial
conditions of x1v1 perturbed in the $z$-direction by $0.1
z_0(\rm{x1v1})$. The evolution of GALI$_2$ with \ej for orbits in the
$-0.3073
\leqq
E_J \leqq -0.2923$ interval is given in
Fig.~\ref{galisd1}. The left column refers to the orbits with the x1v1 initial
conditions perturbed in the $x$-direction, while in the right column to the
orbits perturbed in the $z$-direction. The \ej of each orbit is given in the
lower left corner of the panels.

At \ej=$-0.3073$, the representative of x1v1 is stable and the perturbed by 10\%
in the $x$- and $z$-directions nearby orbits (left and right panels in
Fig.~\ref{galisd1}a) are regular, apparently belonging to a quasiperiodic orbit
trapped around it. The orbits in Fig.~\ref{galisd1}b are at an \ej just beyond
the S$\rightarrow\!\!\Delta$ transition, namely \ej=$-0.3023$, where x1v1 is
already complex unstable. However, GALI$_2$ hardly indicates a chaotic orbit.
Contrarily, its variation points to a regular one. The quasi-regular behaviour
of orbits
close to complex unstable periodic orbits beyond, but close to, the critical \ej
at which the S$\rightarrow\!\!\Delta$ transition occurs, has been formerly
noticed by \citet{pz94} and \citet{kpc11}. Close to the maximum $|\Delta|$, at
\ej=$-0.2973$, the GALI$_2$ index of the perturbed orbit identifies a chaotic
behaviour as it becomes practically zero, reaching values
at the order of computational accuracy, i.e. $\approx
10^{-16}$ (Fig.~\ref{galisd1}c). Interesting is that the same amount of
perturbation at \ej=$-0.2923$, when x1v1 is again stable, gives again chaotic
orbits, as the variation of the GALI$_2$ indices show. This happens because
this perturbation brings the initial conditions of the perturbed orbit, beyond
the volume occupied by the invariant tori around the stable x1v1 at this
\ej\!\!.

\begin{figure}
\centering
\includegraphics[scale=0.4]{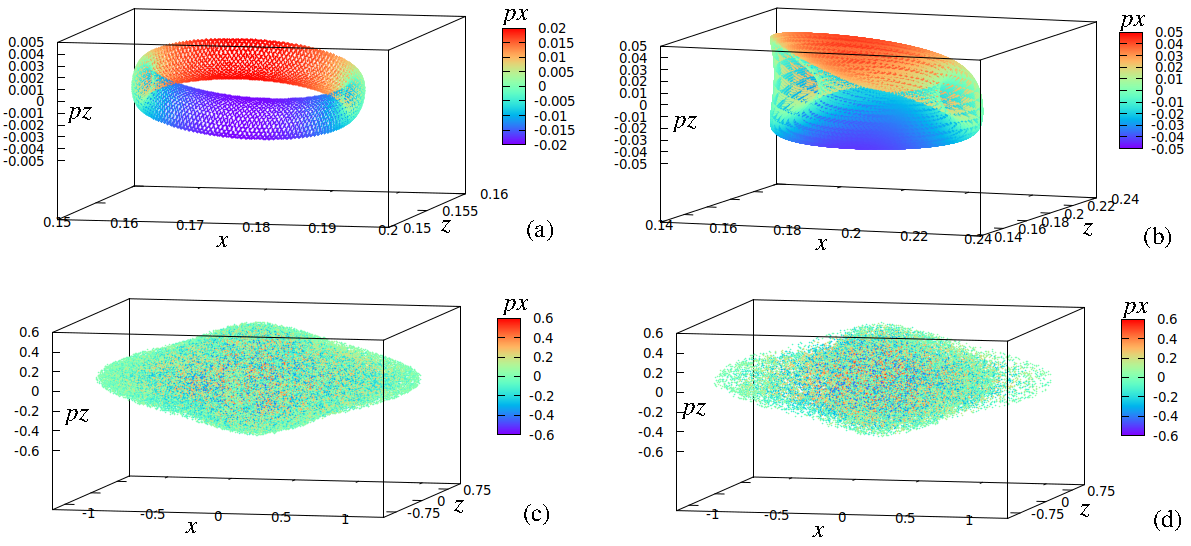}
\caption{Ferrers bar: Poincar\'{e} sections of orbits close to x1v1 periodic
orbits at
\ej$=-0.3073$ (a), $-0.3023$ (b), $-0.2973$ (c) and $-0.2923$ (d). All of them
have initial conditions of x1v1 perturbed in the $x$-direction by 0.1$
x_0(\rm{x1v1})$. The x1v1 family at the corresponding energies is stable
in (a) and (d) and complex
unstable in (b) and (c).}
\label{delta1_10pc}
\end{figure}

To the same conclusions leads also the study of the phase-space structure.
For the visualization of the distribution of the $(x,z,p_{x},p_{z})$
consequents in the four-dimensional (4D) space, we use in
Fig.~\ref{delta1_10pc}, and in
subsequent similar figures in the paper, the method proposed by \citet{pz94}.
Namely, we consider a 3D projection of the orbit and we rotate it, by means of
an appropriate software package, in order to understand whether its consequents
are lying on a specific surface, or if they are scattered in the 3D space. Then,
we colour the
consequents according to the value of the fourth coordinate, using a colour
palette. If the consequents lie on a surface, the colours allow us to discern
between a smooth variation of the shades on this surface, or if we have mixing
of colours. This method led to the association of specific structures in
the
neighbourhood of a periodic orbit, in the 4D space of section, with stability,
as well as with each kind of
instability \citep[for details see][]{kp11, kpc11, kpc13}.

The orbits in Fig.~\ref{delta1_10pc}
correspond to the four orbits in the left column of Fig.~\ref{galisd1}. The
three spatial coordinates used for the presentation are $(x,z,p_{z})$, while the
colour of the points is defined by the value of their $p_{x}$ coordinate,
according to the palette given on the right hand side of each panel. The
integration time of the orbits depicted in Fig.~\ref{delta1_10pc} is much longer
than the 5~Gyr period, we used in the calculation of the GALI$_2$ indices, since
we want to have a clear view of the formed structures. Thus, we continued
integrating the orbits even for times beyond the realistic limits of the
physical system.

In Fig.~\ref{delta1_10pc}a, at \ej$\!=\!-0.3073$, x1v1 is stable and the
consequents of the plotted non-periodic orbit form a toroidal structure with a
smooth colour variation on its surface, according to the palette given at the
right hand side of the panel. Such a structure in the 4D surface of section,
points to a quasi-periodic orbit trapped around a stable periodic orbit
\citep{kp11}. In Fig.~\ref{delta1_10pc}b, at \ej$=-0.3023$, we are beyond the
S$\rightarrow\!\!\Delta$ transition and x1v1 is now complex unstable.
Nevertheless, the consequents of the orbit form again in the $(x,z,p_{z})$
projection a toroidal, albeit more complicated, structure than the one depicted
in Fig.~\ref{delta1_10pc}a. It has also a hole at the center, which is not
discernible in Fig.~\ref{delta1_10pc}b, because we use the same point of view
for all four panels in Fig.~\ref{delta1_10pc}. However, it can be observed e.g.
in the ($x,p_x$) projection. This implies that the orbit with initial
conditions those of the periodic orbit, perturbed in the $x$-direction by 0.1$
x_0(\rm{x1v1})$ may have reached tori around another, stable, periodic orbit,
existing in this phase space region.
The internal architecture of structures in phase
space around complex unstable periodic orbits in Poincar\'{e} cross sections and
their gradual deformation as one parameter of the model, in our case \ej\!\!,
varies, has been investigated in several cases in the past \citep{pcp95, kpc11,
sb21}. For the needs of the present study, it is evident that in
Fig.~\ref{delta1_10pc}c, when we have reached \ej\!$=-0.2973$, close to the
maximum
$|\Delta|$ of the complex unstable region, the points of the perturbed x1v1,
nonperiodic, orbit appear scattered and their colours mixed. This indicates a
chaotic orbit, as also its GALI$_2$ suggests (Fig.~\ref{galisd1}c). Finally, if
we consider an orbit at \ej\!$=-0.2923$, beyond the $\Delta\!\!\rightarrow$S
transition, where x1v1 is again stable and we perturb the periodic orbit by
$0.1x_0\rm(x1v1)$ in the $x$-direction, as in all previous cases of
Fig.~\ref{galisd1}, we encounter a chaotic behaviour (Fig.~\ref{delta1_10pc}d).

\begin{figure}
\centering
\includegraphics[scale=0.4]{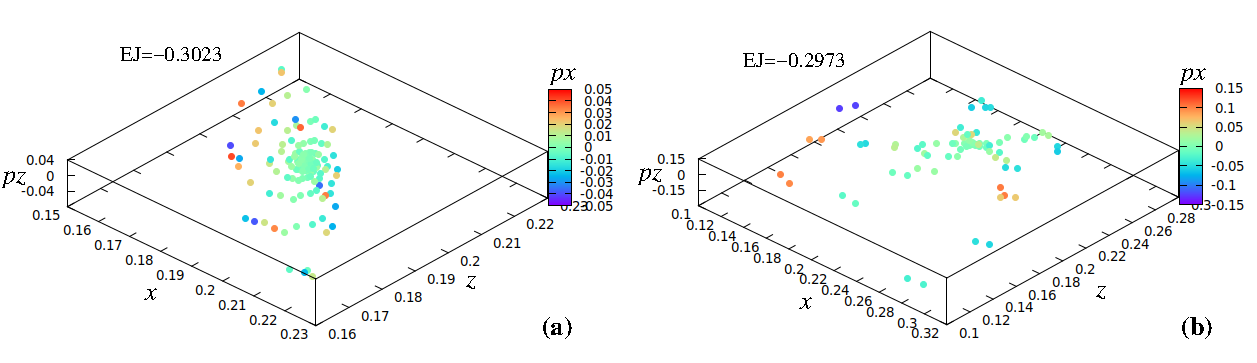}
\caption{Ferrers bar: The first few hundreds of the consequents in the
Poincar\'{e} sections
of two orbits in the immediate neighbourhood of complex unstable x1v1 periodic
orbits, to which we have applied just a tiny perturbation, of the order
of $10^{-9}$ in the $x$ coordinate. In (a) \ej$=-0.3023$, as in
Fig.~\ref{delta1_10pc}b and in (b) \ej=$-0.2973$, as in
Fig.~\ref{delta1_10pc}c.}
\label{d1_2complex}
\end{figure}

It is obvious that, at least in this case, the regular or chaotic character in
the vicinity of the periodic orbit, at the level of a perturbation of 10\% of
just one of the four initial conditions of it, is not associated with the value
of the quantity $\Delta$. Even more, in some cases like the one presented in
Fig.~\ref{galisd1}d and Fig.~\ref{delta1_10pc}d, the orbit is not affected at
all by the presence of the stable periodic orbit x1v1. Apart from the degree of
complexity of the 3D projections of the regular structures around a stable
(Fig.~\ref{delta1_10pc}a) and a complex unstable (Fig.~\ref{delta1_10pc}b)
periodic orbit, the main difference in the structure of phase space between the
two cases is found in the immediate neighbourhood of the periodic orbit, i.e.
within a radius $r$ around it, as $r\rightarrow 0$. Around a stable periodic
orbit we always find toroidal structures, while for tiny perturbations of the
initial conditions of a complex unstable one, the consequents drift away from it
without building a hole. In the case of the complex unstable periodic orbit at
\ej=$-0.3023$, at which the regular structure of Fig.~\ref{delta1_10pc}b also
exists, a perturbation of $10^{-9}$ of its $x$ coordinate leads to an orbit
with the Poincar\'{e} cross section we present in Fig.~\ref{d1_2complex}a. A
spiral pattern around the initial conditions of the periodic orbit, like those
encountered in previous studies \citep{pcp95, kpc11, sb21},  is discernible for
the first 350 consequents (Fig.~\ref{d1_2complex}a). Then, gradually, a regular
structure is formed with increasing integration time. Contrarily, around
the
complex unstable periodic orbit with \ej=$-0.2973$, in the case of the orbit
in Fig.~\ref{delta1_10pc}c, we do not observe a spiral pattern
even for tiny
perturbations. In Fig.~\ref{d1_2complex}b we give the first 200 consequents of
such an orbit. We can only observe that the consequents depart from the
periodic orbit along certain directions.

Beyond the $\Delta\!\!\rightarrow$S transition, in order to find regular orbits
close to the now stable x1v1, i.e. quasi-periodic orbits trapped around it at
\ej=$-0.2923$, we have to reduce the perturbation applied to the $x_0$ initial
condition, to $0.02x_0\rm{(x1v1)}$. The toroidal structure we find in 4D has a
hole, evident in the $(x,p_x)$ projection. If we go back to the
complex
unstable region and we consider a complex unstable periodic orbit at
\ej=$-0.2938$, where $\Delta$ is about the same as $\Delta$ at \ej$=-0.3023$
(Fig.~\ref{ejdel_005a}), we do not find similar phase structures around the
two complex unstable periodic orbits. In the
smaller energy (\ej$=-0.3023$) we have encountered the orbit presented in
Fig.~\ref{delta1_10pc}b, with the GALI$_2$ shown in Fig.~\ref{galisd1}b by
perturbing the $x$ coordinate by 10\%. In the larger energy (\ej=$-0.2938$),
by applying the same relative perturbation we find chaos. In this case,
even
if we reduce the perturbation in the $x$-direction to $0.001x_0\rm{(x1v1)}$, we
find chaotic orbits. There is no symmetry in the
phase space structure in the cases of the two periodic orbits with similar
$\Delta$ (Eq.~\ref{delta}) values. This is better realized close to the
transition
points (S$\rightarrow \Delta$ and $\Delta \rightarrow$ S), where we observe that
around the orbit with the larger \ej\!\! we find more chaos. We also find that
beyond the $\Delta\rightarrow$S transition the volume of regular orbits around
the stable periodic orbits is reduced. In most cases, this asymmetry
reflects
the different landscapes we encounter in the phase space region around the
periodic orbits of a family at different \ej. Thus, by applying similar
perturbations at different \ej we may enter a zone of influence of a
stable periodic orbit or a chaotic sea.
Nevertheless, within a time of interest for the specific physical problem, i.e.
for 5~Gyr, orbits with initial conditions deviating from those of the complex
unstable x1v1 by $0.05x_0\rm{(x1v1)}$ or $ 0.05z_0\rm{(x1v1)}$ are to a large
degree bar-supporting.

\begin{figure}
\centering
\includegraphics[scale=0.3]{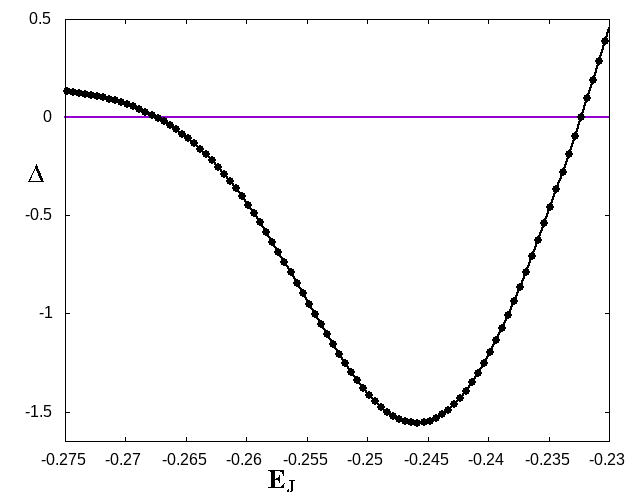}
\caption{Ferrers bar: Variation of $\Delta$ (Eq.~\ref{delta}) for the x1v1
family of periodic
orbits in the
$-0.275 \leq E_J  \leq -0.23$ region. The heavy dots correspond to
calculated periodic orbits. Those with $\Delta < 0$ are complex unstable.}
\label{ejdel2_005a}
\end{figure}
\begin{figure}
\centering
\includegraphics[scale=0.4]{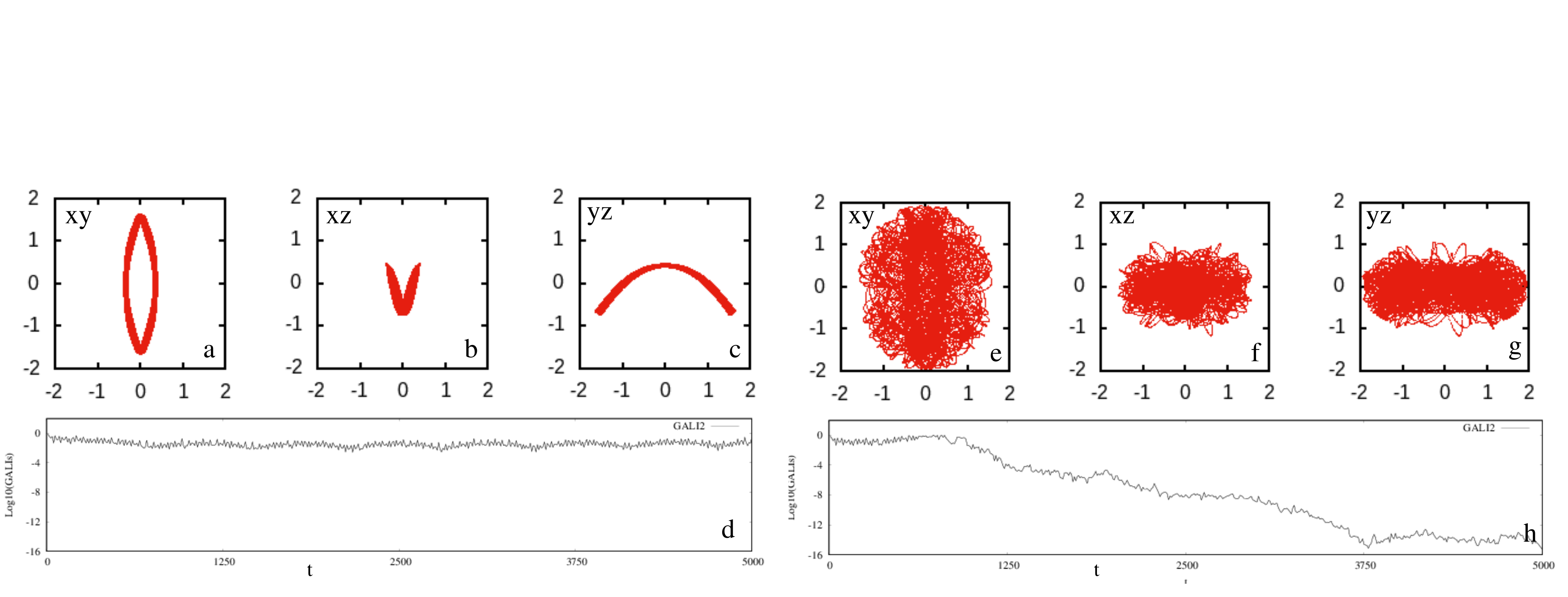}
\caption{Ferrers bar: From (a) to (c) we give the $(x,y)$, $(x,z)$ and $(y,z)$
projections of an x1v1 orbit
with \ej=$-0.27134$,
perturbed by 0.05$x_0(\rm{x1v1})$ in the $x$ coordinate. They point to
a
quasi-periodic orbit trapped around the x1v1 periodic orbit, as its GALI$_2$
variation over a 5~Gyr period indicates in (d). (e) to (g): The corresponding
projections for the x1v1 orbit perturbed by 0.05$z_0(\rm{x1v1})$ in the $z$
coordinate.
The orbit behaves in a chaotic way, after an initial sticky phase, as
its GALI$_2$
variation in (h) indicates.}
\label{troxies9293}
\end{figure}

\vspace{0.5cm}
The second $\Delta$ region in this model, is found for $-0.2673 \lessapprox E_J
\lessapprox -0.2328$, where we have again a S$\rightarrow\!\!\Delta\!\!
\rightarrow$S transition. This time, the complex unstable region extends in a
broader \ej range and the maximum $|\Delta|$ in it is much larger
(Fig.~\ref{ejdel2_005a}).
For orbits at these energies the dynamical time scales are large and to the same
physical time correspond much less consequents. Perturbations of the stable
orbits of the x1v1 family of the order of 0.1$x_0(\rm{x1v1})$ or
0.1$z_0(\rm{x1v1})$ in the $x$- or $z$-direction respectively, bring
always the initial conditions in chaotic regions of phase space. In that sense,
before the second S$\rightarrow\!\!\Delta$ transition of x1v1, at
\ej$\approx-0.2673$, we enter chaotic seas by applying relatively smaller
perturbations than to the initial conditions of the stable periodic orbits of
the family before the first S$\rightarrow\!\!\Delta$ transition of x1v1, at \ej
$\approx -0.3028$. By reducing the perturbations to 0.05$x_0(\rm{x1v1})$,
or 0.05$z_0(\rm{x1v1})$, we find regular, i.e. quasi-periodic, orbits
around stable x1v1 for \ej$<\!-0.2713$. Then, as we approach the critical \ej at
$\approx-0.2673$, the perturbed by 5\% orbits become chaotic, first along the
$z$ direction, while even closer to it we have to reduce the perturbation even
more in order to find close to the periodic orbits regular structures in phase
space. In Fig.~\ref{troxies9293} we give the perturbed by 0.05$
x_0(\rm{x1v1})$ (panels a to d) and 0.05$z_0(\rm{x1v1})$ (panels e to h)
orbits, for \ej=$-0.27134$.
The perturbed in the $x$-direction orbit has a typical quasi-periodic morphology
as its $(x,y)$, $(x,z)$ and $(y,z)$ projections, in Fig.~\ref{troxies9293}a, b
and c respectively, show. The regular nature of the orbit is in agreement with
the variation of its GALI$_2$ index in Fig.~\ref{troxies9293}d. Contrarily, the
orbit perturbed in the $z$-direction (Fig.~\ref{troxies9293}e,f,g) is chaotic.
The GALI$_2$ index (Fig.~\ref{troxies9293}h) shows that after an initial sticky
phase, the orbit has a chaotic behaviour. In order to find regular orbits when
we perturb x1v1 in $z$ at this and larger \ej\!\!, before the
S$\rightarrow\!\!\Delta$ transition, we have to impose perturbations of the
order of $10^{-3}z_0(\rm{x1v1})$ or smaller.

\begin{figure}
\centering
\includegraphics[scale=0.4]{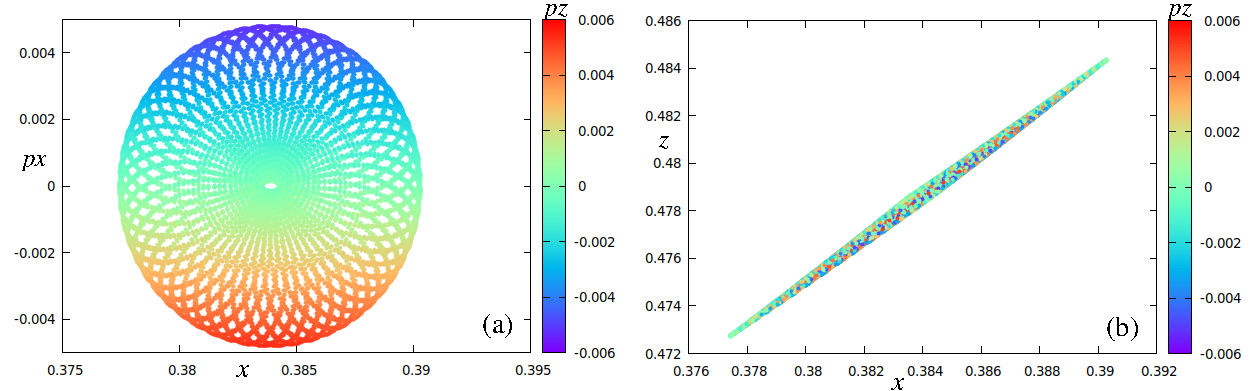}
\caption{Ferrers bar: The $(x,p_x)$ (a) and $(x,z)$ (b) projections of the
orbit with
initial conditions those of the stable x1v1 at
\ej=$-0.26753617$, perturbed by $0.001 x_0\rm{(x1v1)}$,
form a very thin toroidal - disky, morphology. The consequents are coloured
according to their $p_z$ values. The energy of the orbit is very close to the
S$\rightarrow\!\!\Delta$ transition. For long integration times, all holes but
the central in (a) are eventually filled.}
\label{d2lasts}
\end{figure}

As we approach the critical \ej\!\!, where we have the S$\rightarrow\!\!\Delta$
transition, the volume of phase space with regular orbits around a stable x1v1,
shrinks. In parallel, there is an evolution of the morphology of tori
structures, e.g. like the one given in Fig.~\ref{delta1_10pc}a, towards a disky
configuration. Nevertheless, we find a hole in the center of these structures.
For example, if we perturb the initial conditions of a stable x1v1 orbit very
close to the $\Delta$ region, at \ej=$-0.26753617$, by $0.001
x_0\rm{(x1v1)}$, we find the orbit depicted in Fig.~\ref{d2lasts}. We give the
$(x,p_x)$ and $(x,z)$ projections, in Fig.~\ref{d2lasts}a and
Fig.~\ref{d2lasts}b respectively, with the consequents coloured according to
their $p_z$ values.

\begin{figure}
\centering
\includegraphics[scale=0.4]{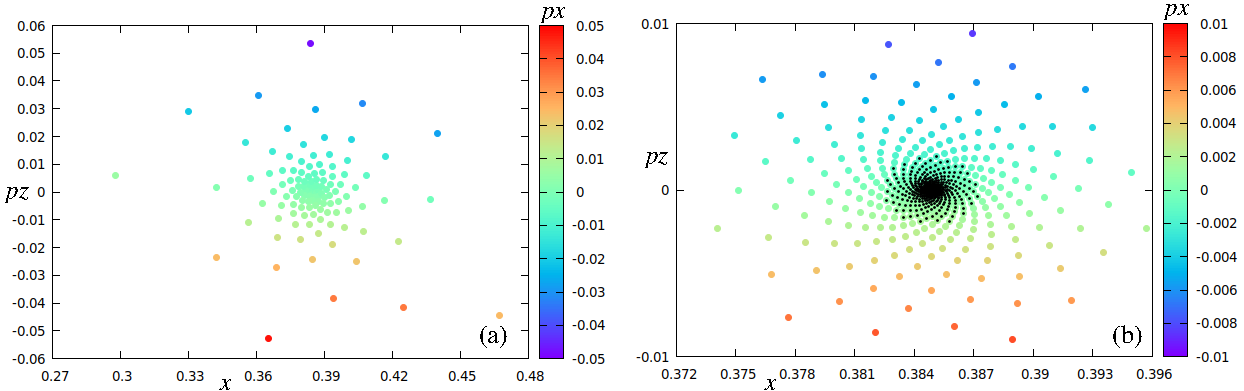}
\caption{Ferrers bar: Perturbed x1v1 orbits at \ej=$-0.26743617$, just after
the
S$\rightarrow\!\!\Delta$ transition. (a) An orbit perturbed by $
0.001 x_0\rm{(x1v1)}$. (b) An orbit perturbed by $10^{-8} x_0\rm{(x1v1)}$. In
(a) there
are 120 consequents plotted, while in (b) 1800, the first 1600 of which are
marked with black dots. For larger integration times both orbits become
chaotic.}
\label{secd1std}
\end{figure}

Beyond the transition, in the immediate neighbourhood of the x1v1 orbits, which
are now complex unstable, we encounter the known arrangement of the consequents
in a spiral lay out \citep{pcp95, kpc11, sb21}, as in the case for
\ej=$-0.26743617$, which we present in Fig.~\ref{secd1std}.
In Fig.~\ref{secd1std}a the x1v1 orbit is perturbed by $0.001
x_0(\rm{x1v1})$. We observe the first 120 consequents, which are organized in a
multi-spiral pattern with smooth colour variation along its arms. In this case,
we give the $(x,p_z)$ projection, in which the points are coloured according to
the value of their $p_x$ coordinate. If we continue integrating the orbit, the
consequents will build a cloud of points with mixed colours, namely the orbit
will behave in a chaotic way. If we consider a tiny perturbation $
10^{-8} x_0(\rm{x1v1})$, we find the corresponding representation of the
Poincar\'{e} surface of section, which is  given in Fig.~\ref{secd1std}b. The
organization of the consequents in the depicted multi-spiral pattern lasts for
about 1800 intersections, the 1600 of which are marked with black dots.
This orbit, for larger integration times behaves also as a chaotic one.

We underline that in both cases the ``regular'' period of the orbits is much
longer than the 5~Gyr time interval we are interested in, for finding
bar-supporting orbits. However, as regards the properties of the dynamical
system we study, we remark that the transition to chaos is more abrupt in the
second than in the first case of the S$\rightarrow\!\!\Delta$ transitions we
discussed. In neither case presented in Fig.~\ref{secd1std} the long time
integration results to the formation of invariant structures around the complex
unstable periodic orbit, such as those encountered in \citet{pf85a} or
\citet{kpc11}. Having orbits, which behave initially as regular and later as
chaotic, we can characterize them as sticky \citep{ch10}. For \ej\!\!'s away
from the critical one, at which we have the second S$\rightarrow\!\!\Delta$
transition, we can hardly trace a spiral pattern in the surfaces of section of
orbits close to x1v1, even if we apply very small perturbations.


The determination of the volume of phase space around a complex unstable
periodic orbit, where we can find structure-supporting orbits, is a heavy task.
Given that even small perturbations may well bring the initial conditions of
the perturbed orbit to zones of influence of other orbital families, not
necessarily simple periodic, it is not always clear to which degree the presence
of a complex unstable periodic orbit is associated with the level of chaoticity
of a nearby orbit. This is also indicated by the variation of its
GALI$_2$ index.

\begin{figure}
\centering
\includegraphics[scale=0.37]{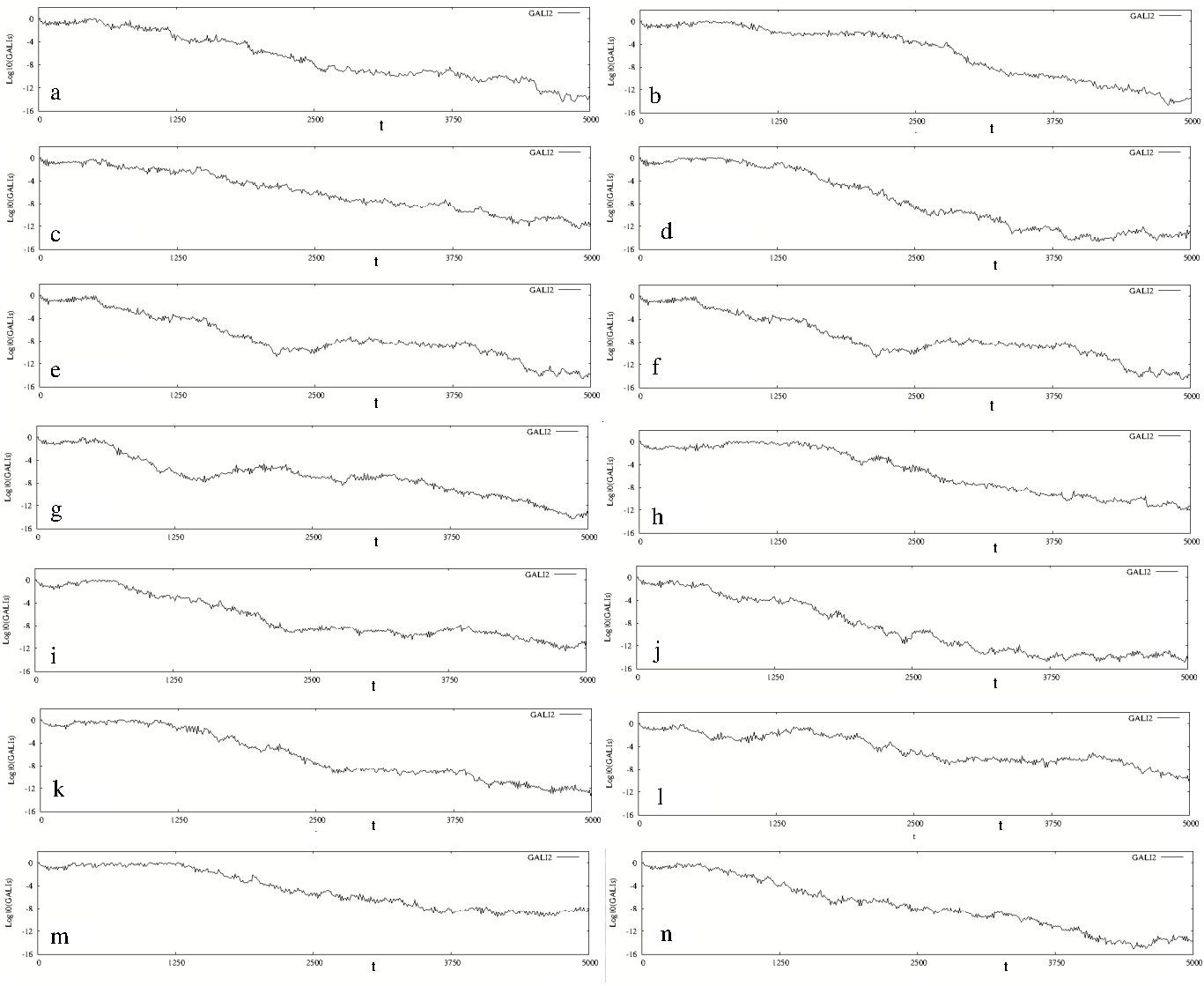}
\caption{Ferrers bar: GALI$_2$ variation of orbits in the neighbourhood of
complex unstable x1v1 orbits, perturbed by $0.05 x_0\rm{(x1v1)}$ (left column)
and by $ 0.05 z_0\rm{(x1v1)}$ (right column). The variation of $\Delta$ of the
corresponding periodic orbits can be deduced from Fig.~\ref{ejdel2_005a} (the
\ej's are given in the text). The panels closest to the maximum $|\Delta|$ are
in the fourth row, i.e. in panels (g) and (h). In general the GALI$_2$ variation
can be affected by the presence of other nearby families existing in phase space
close to the periodic orbit.}
\label{g2d2}
\end{figure}

For instance, in Fig.~\ref{g2d2} we consider orbits in the neighbourhood of
seven
complex unstable periodic orbits in the \ej interval $-0.275 \lessapprox E_J
\lessapprox -0.23$ (Fig.~\ref{ejdel2_005a}) with initial conditions close to
those of x1v1, but with $x_0=1.05 x_0(\rm x1v1)$ or with $z_0= 1.05
z_0(\rm x1v1)$, and we plot the variation of their GALI$_2$ indicators within a
5~Gyr period. The panels on the left hand side correspond to the orbits
perturbed by $\Delta x$, while on the right hand side to the orbits perturbed by
$\Delta z$. The Jacobi constants and the $\Delta$ value of the corresponding
complex unstable x1v1 periodic orbit are: In (a) and (b) \ej=$-0.26534465$ and
$\Delta\approx-0.083$, in (c) and (d) $-0.25484465$ and $-0.951$ respectively,
in (e) and (f) $-0.25334465$ and $-1.108$, in (g) and (h) $-0.24584465$ and
$-1.557$, close to the largest $|\Delta|$, in (i) and (j) $-0.23984465$ and
$-1.196$, in (k) and (l) $-0.23684465$ and $-0.786$, and finally in (m) and (n)
$-0.23384465$ and $-0.276$.

In the left column of Fig.~\ref{g2d2}, we observe that there is always an almost
horizontal part of the curves with the GALI$_2$ variation, which appears at the
left side of each panel. This part corresponds to times of the order of 1~Gyr or
less. For the dynamical time scales of these orbits, at the \ej we
consider them, within this
period we have only a few consequents, less than 20, which depart from the
periodic orbit forming in general a spiral pattern, before they start behaving
in a chaotic way. In panel (g), where we have a perturbed x1v1 orbit in the
$x$-direction, we are closest to the maximum $|\Delta|$ of the region we study.
We observe that the horizontal branch of its GALI$_2$ variation is, together
with the one in panel (e), one of the shortest. However, for larger times in
Fig.~\ref{g2d2}g, there is a second plateau, before the curve starts decreasing
monotonically. Such variations make it even more difficult to link the values of
$\Delta$ (Eq.~\ref{delta}) with the degree of chaoticity to the phase space
around a complex unstable periodic orbit.

\begin{figure}
\centering
\includegraphics[scale=0.43]{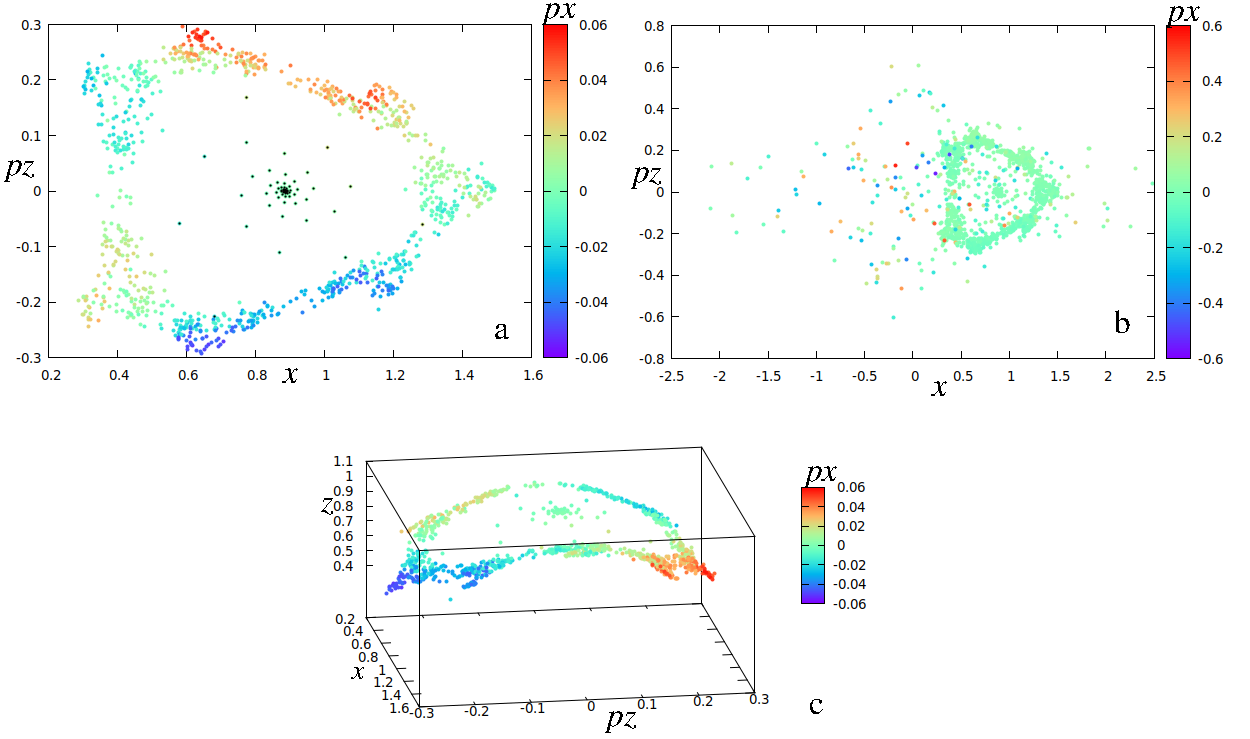}
\caption{Ferrers bar: An orbit close to the x1v1 complex unstable periodic
orbit, at \ej\!\!= $-0.23283617$, near the $\Delta\!\!\rightarrow$S transition.
The initial conditions of the orbit differ from those of the periodic one by
$10^{-8}x_0$ in the $x$ coordinate. In (a) and (b) we give the $(x,p_z)$
projection, with the consequents coloured according to their $p_x$ values, while
in (c) the $(x,p_z,z)$ projection, coloured again according to the $p_x$ values.
In (a) and (c) we consider 1200 intersections with the $y$=0 plane, while in (b)
1600. The orbit drifts to a chaotic sea after forming first a spiral pattern and
then remaining sticky in a zone around the periodic orbit.}
\label{cen2328}
\end{figure}

A characteristic example of a complicated landscape of the phase space in the
neighbourhood of a complex unstable periodic orbit is given in
Fig.~\ref{cen2328}. We present the $(x,z,p_x,p_z)$ Poincar\'{e} section of an
orbit very close to the $\Delta\!\! \rightarrow$S transition, at the right hand
side of the $\Delta$ region in Fig.~\ref{ejdel2_005a}, at \ej\!\!=
$-0.23283617$,
where x1v1 has $\Delta=-0.092$.
We consider the periodic orbit and apply a tiny perturbation in its initial
$x_0$ condition, namely $10^{-8}x_0(\rm{x1v1})$. The first 226 consequents of
the orbit form a usual spiral pattern (central region of Fig.~\ref{cen2328}a),
as they deviate away from the complex unstable x1v1. However, the breaking of
the spiral pattern is not followed by a diffusion in a chaotic domain, but by
the sticking of the orbit in a weakly chaotic zone surrounding a chain of
stability islands. In Fig.~\ref{cen2328}a, we give the first 1200 consequents in
the $(x,p_z)$ projection, coloured according to their $p_x$ values. The first
226, building a 3-armed spiral pattern, are emphasized with black points. In
Fig.~\ref{cen2328}b we give the first 1600 consequents and we observe how they
diffuse in a broader chaotic sea. In Fig.~\ref{cen2328}c we present again the
first 1200 consequents, but using the 3D $(x,p_z,z)$ projection, also coloured
according to their $p_x$ values. We realize that the consequents are practically
on a warped-disky surface, reminiscent of the shape of the disky confined tori
\citep{kpc11}.

We reach similar conclusions by studying perturbations in the $z$-direction. In
Fig.~\ref{g2d2}, the right hand column with the GALI$_2$ indices
refers to orbits with initial conditions close to the periodic orbits,
perturbed in the $z$ direction by $0.05 z_0\rm{(x1v1)}$. This time,
from top to
bottom, the horizontal part of the GALI$_2$ curves initially is reduced with
increasing $|\Delta|$. However, in panel (h), close to the x1v1 orbit
with the maximum $|\Delta|$, the perturbed by 5\% in the $z$-direction orbit
shows a more extended horizontal part, as does the orbit in panel (l). Such
variations are again due to the presence of the orbits of other families in the
neighbourhood of the periodic orbit we study.

\begin{figure}
\centering
\includegraphics[scale=0.27]{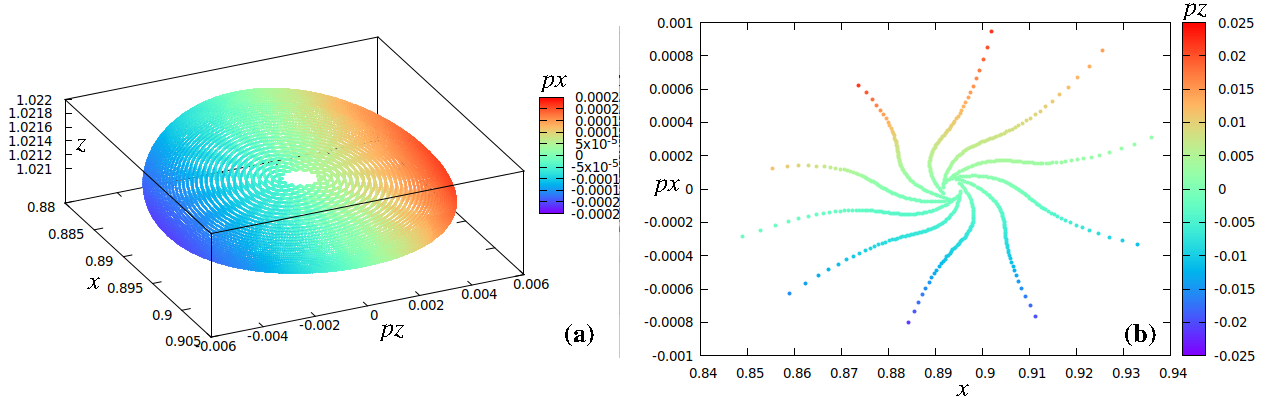}
\caption{Ferrers bar: (a) The tiny torus of the quasiperiodic orbit with
initial conditions
of x1v1, perturbed by $0.0015 x_0(\rm{x1v1})$, for
\ej\!\!=$-0.23234$. We use the $(x,p_z,z)$ spatial projection and $p_x$ for
giving
colour to the consequents. (b) The first 600 consequents of another orbit,
this time with a
perturbation $0.002 x_0(\rm{x1v1})$, at the same \ej\!\!, in the
$(x,p_x)$ projection, coloured according to their $p_z$ value. They form a
structure reminiscent of the spirals close to complex unstable periodic
orbits. For longer integration times this orbit diffuses in the available phase
space.}
\label{stjustbeyond}
\end{figure}

Restoration of stability, for \ej$ > -0.2328$, has also a gradual character.
Just beyond the $\Delta\!\!\rightarrow$S transition, the ``range of influence''
of the stable periodic orbit is small. For \ej\!\!=$-0.23234$, the
tolerance of the perturbation of the $x_0$ initial condition, so that we find
quasiperiodic orbits on tori with a smooth colour variation on them, is just
$\Delta x \approx 0.0015 x_0(\rm{x1v1})$. This orbit can be seen in
Fig.~\ref{stjustbeyond}a. For larger perturbations of $x_0$ the orbits are
chaotic, with an initial sticky phase appearing up to a perturbation of $\Delta
x \approx 0.05 x_0(\rm{x1v1})$. In Fig.~\ref{stjustbeyond}b, we give the first
600 consequents of the orbit, for which the perturbation of x1v1 is $
0.002 x_0(\rm{x1v1})$. During this period, the structure of phase space around
the stable periodic orbit resembles the spiralling observed around a
complex unstable one. For larger integration time a chaotic cloud is formed,
similar to those depicted in Fig.~\ref{delta1_10pc}c,d. Away from the transition
to stability region, the phase space structure around the stable x1v1 orbits is
characterized by stability islands of considerable sizes. For example, for
\ej=$-0.22934$, if we perturb again the periodic orbit in $x$, we
find tori of quasi-periodic orbits for perturbations up to about $0.5
x_0(\rm{x1v1})$.

\subsection{Complex Unstable regions in PERLAS spirals}
\label{perlas}
In the PERLAS case the perturbative term is in the form of a spiral potential,
in which the mass of the spiral ($M_{s}$), over the mass of the disc ($M_{d}$)
is $M_{s}/M_{d}=0.04$ \citep[model M4 in][]{chvv19}. The family of x1v1 periodic
orbits is introduced in the same way as in the rotating Ferrers bar,
namely as a
bifurcation of the central family x1, at the vertical 2:1 resonance. The
projections of the orbits of this family on the equatorial plane are
elliptical-like. However, in the PERLAS potential, they are not aligned along an
axis as in the case of a bar. Their orientation changes in such a way, as to
support a bisymmetric set of logarithmic spiral arms \citep{chvv19}.

In the specific PERLAS model we study, the evolution of the stability of this
family with \ej is also qualitative similar with that of the Ferrers bar we
studied in the previous subsection (\ref{deltafer}). Namely, we find two
S$\rightarrow\!\!\Delta\!\!\rightarrow$S transitions, for $-1354.3  \lessapprox$
\ej
$\lessapprox -1345.8$ and $-1208.7 \lessapprox$ \ej $\lessapprox -1131.6$, in
the units we use for this model
\citep[see figure 6 in][]{chvv19}. Taking into account that the center of the
system is at \ej $\approx -1580$ and the Lagrangian point L4 at corotation, at
\ej $\approx -1038$, the first complex unstable region is tiny in the energy
range in which the families of the x1-tree extend. The quantity $\Delta$
(Eq.~\ref{delta}) in
$-1354.3 \lessapprox$ \ej $\lessapprox -1345.8$ has a variation similar to
those in the complex
unstable regions we presented in Fig.~\ref{ejdel_005a} and
Fig.~\ref{ejdel2_005a} for the Ferrers bar model, with a maximum
$|\Delta|\approx 0.029$. In a 5~Gyr period, all orbits with initial conditions
those of the complex unstable periodic orbits perturbed in the $x$-direction by
$0.1 x_0(\rm{x1v1})$ or in the $z$-direction by $0.1
z_0(\rm{x1v1})$ behave apparently as regular and support the imposed 2-armed
spiral pattern. We calculated their GALI$_2$ indices and
we found variations indicating a regular behaviour.

The second complex unstable region ($-1208.7 \lessapprox$ \ej
$\lessapprox -1131.6$) is
quite broad
and the variation of $\Delta$ (Eq.~\ref{delta}), which is again U-shaped as in
all previous cases,
has a minimum $\Delta=-4.9$ at \ej$\approx -1165$. Let us first describe the
evolution of structures in phase space close to the S$\rightarrow\!\!\Delta$
transition point at \ej = $-1208.7$. We study it by applying radial
perturbations to the $x$ coordinate of the initial conditions of the x1v1
periodic orbit. For the sake of brevity in the case of the spiral PERLAS
potential we will use for the presentation of our results mainly radial
perturbations. We do so, because the problem of the orbital support of
a
galactic
grand-design spiral pattern should be considered in a first approximation as a
problem of finding perturbed orbits practically on the equatorial plane of the
model.

\begin{figure}
\centering
\includegraphics[scale=0.285]{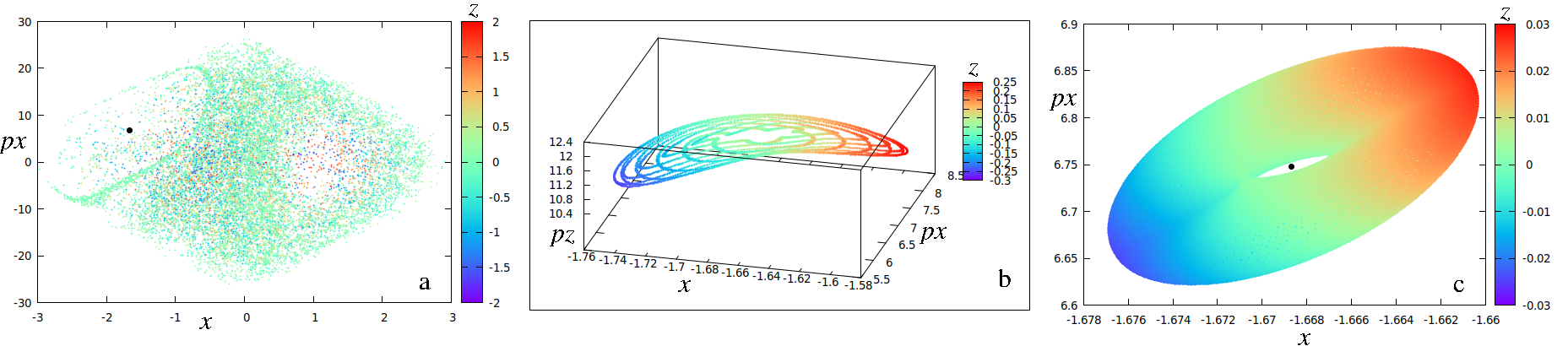} \caption{PERLAS potential: The
cross sections of three orbits in the neighbourhood of the stable x1v1 periodic
orbit at \ej = $-1210.228$, just before the S$\rightarrow\!\!\Delta$ transition.
All of them are perturbations of the periodic orbit in the $x$ coordinate. In
(a) by $x = 0.03x_0(\rm{x1v1})$, in (b) by $0.01x_0(\rm{x1v1})$ and in (c) by
$0.001x_0(\rm{x1v1})$. The area occupied by regular orbits around x1v1 is small.
Black dots in (a) and (c) point to the location of the periodic orbit.}
\label{s1210}
\end{figure}

Close to the transition point, at \ej$= -1210.228$, x1v1 is still
stable.
However, we find again that the extent of the region within which we find
quasiperiodic orbits around x1v1, has been considerably reduced. Already a
perturbation by
$0.03x_0(\rm{x1v1})$ corresponds to a chaotic orbit, which visits
all available phase space if integrated for long time, as we can see in
Fig.~\ref{s1210}a. The location of x1v1 in Fig.~\ref{s1210}a is indicated with a
heavy black dot at the left part of the figure. It is located at
negative $x$, due to the way we define the
Poincar\'{e} section in a clockwise rotating system. For a perturbation $
0.01x_0(\rm{x1v1})$ we find a regular orbit, which is confined in a
thin, warped, disky structure, which is not filled even after $10^4$
intersections (Fig.~\ref{s1210}b). Only for perturbations of the order
of $
0.001x_0(\rm{x1v1})$, we encounter the known, characteristic
structure of a torus in 3D projections with smooth colour variation in the
fourth coordinate (Fig.~\ref{s1210}c).

\begin{figure}
\centering
\includegraphics[scale=0.4]{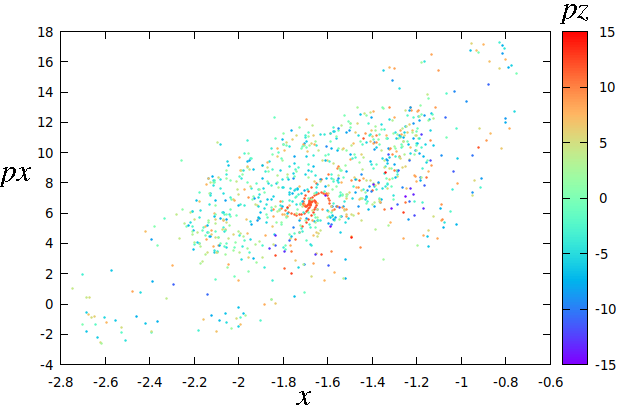}
\caption{PERLAS potential: The $(x,p_x)$ projection of the cross section of an
orbit in the
neighbourhood of the complex unstable x1v1 with \ej = $-1208.228$. The
consequents are coloured according to their $p_z$ values. The orbit has the
initial conditions of x1v1 perturbed by $0.001x_0(\rm{x1v1})$. A
spiral
pattern formed by the first 67 consequents is discernible in the middle of the
figure.}
\label{c1208}
\end{figure}

At a slightly larger \ej\!\!, for \ej = $-1208.228$, the periodic orbit has
become complex unstable, still being  close to the S$\rightarrow\!\!\Delta$
transition. We find that only tiny perturbations of the $x_0$ initial condition
of the periodic orbit result to the formation of spiral patterns around the
periodic orbit in phase space \citep{pcp95, kpc11, sb21}. Nevertheless, even in
these cases, the consequents eventually diffuse in phase space. A characteristic
example is given in Fig.~\ref{c1208}, where we present the cross section of the
orbit with initial conditions those of the periodic orbit x1v1 perturbed by
$0.001x_0(\rm{x1v1})$.

\begin{figure}
\centering
\includegraphics[scale=0.4]{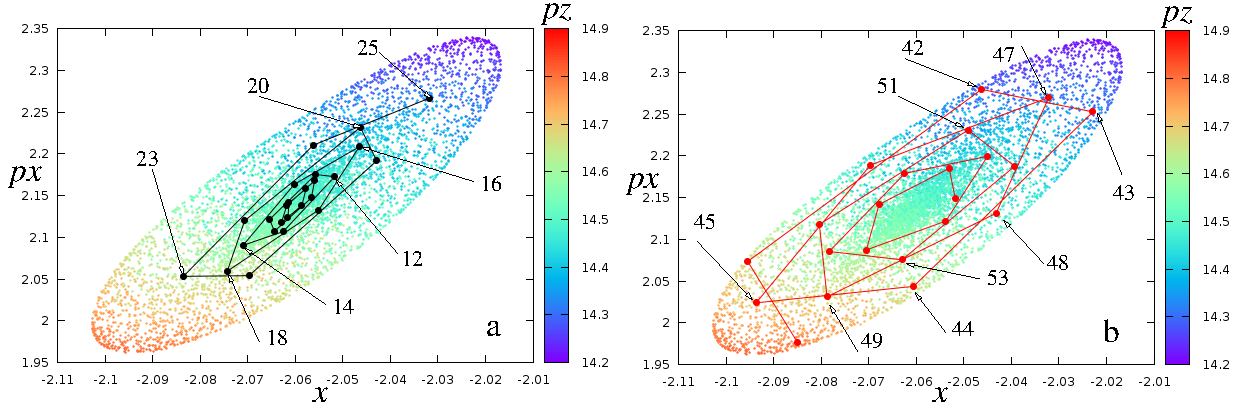}
\caption{PERLAS potential: The disky structure with the smooth colour
variation
along its surface, formed by the consequents of the perturbed by  $\Delta x =
0.001x_0$ x1v1 orbit, at \ej = $= -1132.228$. We give the  $(x,p_x)$ projection
and we colour the points according to their $p_z$ values (colour bar at the
right hand side of the panels). The first 25 consequents are plotted in (a)
with black dots and are connected with black lines and the consequents from the
40th to the 61st one are plotted with red dots and lines in (b). The disky
structure is
formed
by the cosequents following successive cycles of spiralling outwards as in (a)
and then spiralling inwards as in (b). We indicate with numbers, and arrows
pointing
to them, several consequents in both panels in order to understand the
out- and inwards spiralling of the points.}
\label{c1132}
\end{figure}

As \ej increases, the phase space close to the complex unstable x1v1 periodic
orbits becomes practically chaotic. The number of consequents arranged in a
spiral pattern when we integrate orbits close to the periodic one, decreases. We
can say that most complex unstable periodic orbits in the range $-1208.7 <$ \ej
< $-1131.6$ are found embedded in chaotic seas. The situation changes again as
we approach the $\Delta\!\!\rightarrow$S transition, for \ej $\approx -1131.6$.
For example, by considering perturbations $0.001x_0$ to the initial
conditions of x1v1, we find invariant structures around the complex unstable
periodic orbits for \ej $\gtrapprox -1134$. Their appearance is preceded by the
presence of consequences confined for a few hundreds of intersections in an
almost disky structure, before they eventually diffuse in phase space. Very
close to the transition point, at \ej $= -1132.228$, we find in the
neighbourhood of the periodic orbit the known wavy, disky structure in the 3D
projection of the space of section, with a smooth colour variation across it,
representing the fourth dimension \citep{kpc11}. A difference with previous
cases is that the underlying spiral pattern followed by the consequents as they
fill the area of the disky structure is one-armed. This is described in
Fig.~\ref{c1132} for the x1v1 orbit perturbed by $0.001x_0(\rm{x1v1})$. The
cross section is given in the $(x,p_x)$ projection, while the colour of the
points corresponds to their $p_z$ values. The consequents follow first
an
one-armed spiral from the center to the outer boundary of the disky structure
and then continue spiralling inwards. This cycle is repeated until the surface
of
the disky structure is covered with points. We show this by plotting the first
25 consequents of the orbit in Fig.~\ref{c1132}a with heavy black dots and
connecting them with straight lines, and then in Fig.~\ref{c1132}b, by
plotting with red dots and lines the consequents from the 40th to the 61st one.
We indicate with numbers some consequents in both panels, in order to
facilitate
understanding that the points follow spiral patterns. The black consequents
follow a spiral outwards, while the red ones a spiral inwards. Following these
first 61 successive intersections of the orbit with the space of section, one
can appreciate the pattern followed by the consequents in forming the
underlying disky structure.

Also in this model, the phase space structure in the neighbourhood of the orbits
with the same $\Delta$ values in the descending and ascending parts of the
($\Delta$,
\ej\!\!) curve is not the same. We can draw only the general conclusion that
regular structures are found only close to the stability transition points.

The evolution of the phase space beyond the $\Delta\!\!\rightarrow$S transition
has also a gradual character. Just beyond the critical value (\ej$\approx
-1131.6$), at \ej = $-1131.228$, orbits with $\Delta x \geqq 0.1x_0$
perturbations of the initial conditions of x1v1 are chaotic. Only for smaller
perturbations of the $x$ initial condition we find quasi-periodic orbits. The
extent of the zone occupied by regular orbits around stable x1v1 orbits
increases, as in the cases we studied in the Ferrers bar potential, for larger
\ej\!\!'s.


\subsubsection{Practical implications}
\label{dwt}
\begin{figure}
\centering
\includegraphics[scale=0.75]{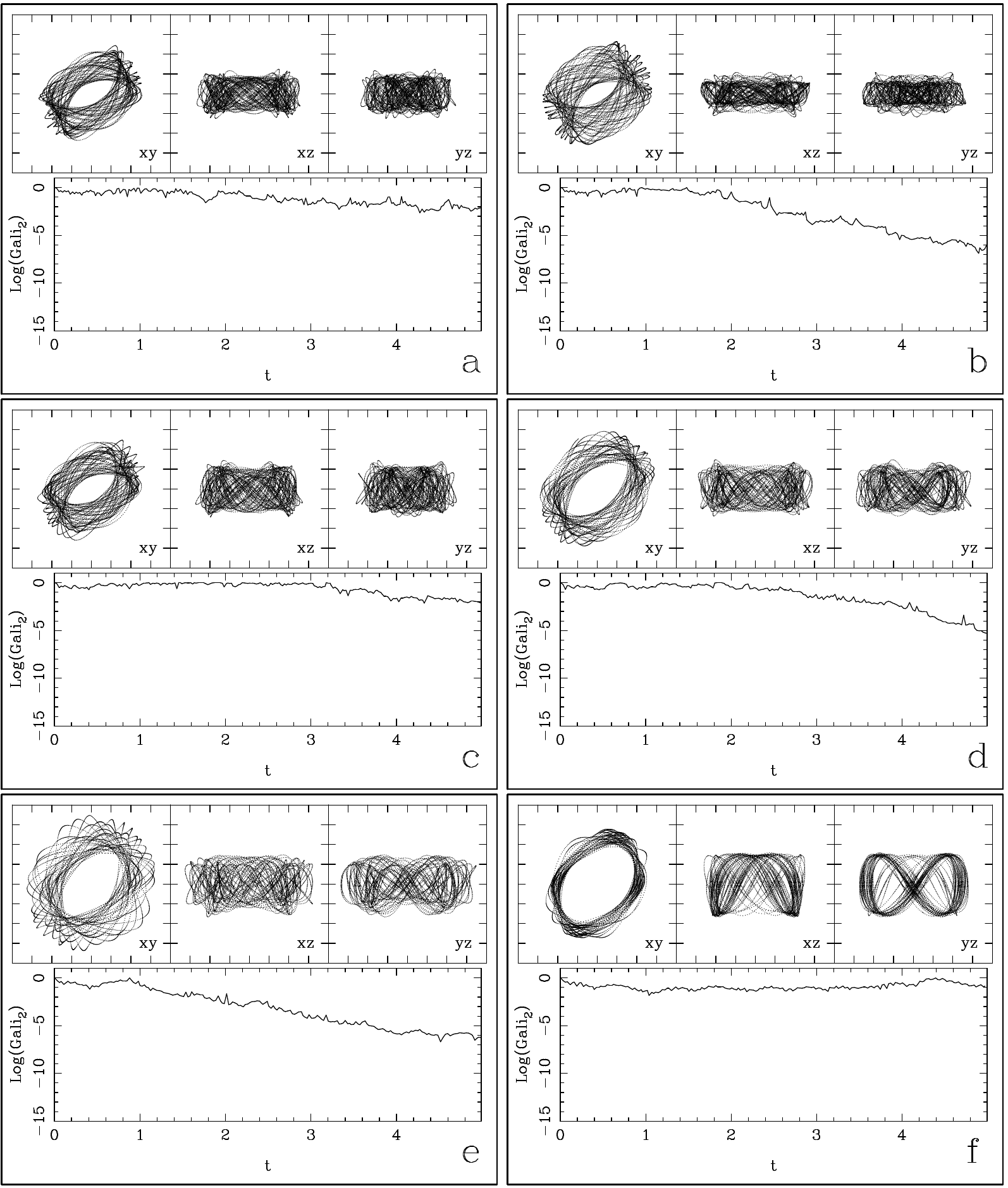}
\caption{PERLAS potential: Perturbed by 10\% in the x coordinate x1v1 orbits.
For each one of them
we present the three projections $(x,y)$, $(x,z)$, $(y,z)$ and the evolution of
GALI$_2$ during a 5~Gyr period. They are at \ej $-1209.228$ (a), $-1208.228$
(b), $-1191.228$ (c), $-1166.228$ (d), $-1132.228$ (e) and $-1131.228 $
(f). The corresponding periodic orbits in (a) and (f) are stable, while in all
other cases complex unstable. All depicted orbits reinforce at some degree the
spiral pattern.}
\label{perlasorb}
\end{figure}

Besides the knowledge of the long-term evolution of the phase space structure in
the neighbourhood of complex unstable periodic orbits, of special importance for
Galactic Dynamics is the behaviour of the orbits during the time within which a
spiral pattern is expected to survive. An upper limit for this can be considered
a 5~Gyr period \citep{sllwd11,d14}. During this time interval, all orbits in the
first complex unstable region of x1v1, for $-1354.3 \lessapprox$ \ej
$\lessapprox -1345.8$, with
initial conditions those of the periodic orbit perturbed by $0.1
x_0(\rm{x1v1})$, can hardly be distinguished from quasi-periodic orbits. During
the same time interval, the perturbed in the same way x1v1 orbits in the second
complex unstable region, $-1208.7 \lessapprox$ \ej $\lessapprox -1131.6$,
evolve as shown in
Fig.~\ref{perlasorb}. There we present 6 orbits, the \ej of which and the
$\Delta$
values of the corresponding orbits successively are: \ej= $-1209.228 \;(\Delta =
0.155)\; \rm{(a)}, -1208.228 \;(\Delta = -0.006)\; \rm{(b)}, -1191.228
\;(\Delta = -2.821)\; \rm{(c)}, -1166.228 \;(\Delta = -4.858)\; \rm{(d)},
-1132.228 \;(\Delta = 0.131)$ (e) and $-1131.228 \; (\Delta = 0.157)$ (f). For
each one of them we give the three projections, $(x,y), (x,z)$ and $(y,z)$ and
below them the evolution of their GALI$_2$ index during a 5~Gyr period.

The periodic orbits existing at the \ej\!'s of the non-periodic orbits depicted
in
Fig.~\ref{perlasorb}a and Fig.~\ref{perlasorb}f are stable, while all other
cases
(Fig.~\ref{perlasorb}b,c,d,e) are complex unstable. Both the morphology and the
variation of the GALI$_2$ indicators of the two orbits in the neighbourhood of
the stable periodic orbits indicate a sticky behaviour. We also observe that the
orbits close to the complex unstable periodic ones never become strongly
chaotic. However, although the perturbation of one of the initial conditions of
the periodic orbit by 10\% leads to realistic initial conditions of an orbit
potentially supporting the spiral structure, it is not necessarily associated
with the immediate environment of the complex unstable periodic orbit. Such a
perturbed orbit may belong to an invariant torus around another, stable,
periodic orbit existing in the region, or it may become an orbit trapped in a
nearby sticky zone. This means that we encounter a situation similar to the
perturbed periodic orbits of the Ferrers bar.

\begin{figure}
\centering
\includegraphics[scale=0.5]{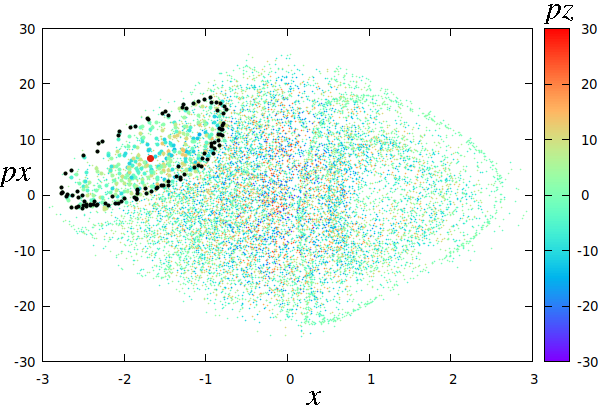}
\caption{PERLAS potential: The $(x,p_x)$ projection of the Poincar\'{e} surface
of section of the
orbit in Fig.~\ref{perlasorb}b. The consequents are coloured according to their
$p_z$ coordinate. During a 5~Gyr period the consequents are found confined
within a ring-like structure delimited by the consequents of the orbit during
the
time interval 3.3 to 4.5~Gyr. This trapping of the consequents in a specific
region of phase space allows the orbit to be spiral-supporting during this
period.}
\label{sos1208}
\end{figure}

For example, the consequents of the orbit in
Fig.~\ref{perlasorb}b, in its $(x,p_x)$ projection of the Poincar\'{e} surface
of section during the 5~Gyr
period, are stuck in the region delimited by the heavy black dots in
Fig.~\ref{sos1208}. These latter, are the consequents of the orbit during the
time interval 3.3 to 4.5~Gyr, which appear stuck along this ring. As the
variation
of the GALI$_2$ of the orbit in Fig.~\ref{perlasorb}b indicates, the orbit is
weakly chaotic, but it does not diffuse in phase space. This secures for this
period the confinement of the orbit in the $(x,y)$ projection on the equatorial
plane in an annular region, which retains the orientation of the x1v1 periodic
orbit with the same \ej\!\!.
This is a useful result, since x1v1 participates in the reinforcement of a
bisymmetric, three dimensional, spiral pattern by means of the mechanism of
``precessing ellipses'' \citep{kal73}, as shown by \citet{chvv19}. The
importance of orbits, which remain encapsulated in regions of phase space for
significant time intervals has been underlined in studies by \citet{muz17,
muz18}.
Our analysis
leads us to the conclusion that even in the complex unstable parts of a family
there are
non-periodic orbits which may contribute to the reinforcement of the spiral
pattern for considerable time intervals. If we continue integrating the orbit
for longer times we find that it
diffuses visiting all available phase space (Fig.~\ref{sos1208}). However, this
happens for non-realistic time scales, of the order of several Hubble times.

\section{Conclusions}
\label{concl}
We have investigated the phase space in the neighbourhood of complex unstable
periodic orbits in two galactic type models, that support structures similar to
those observed in disc galaxies. The models rotate with a constant pattern
speed. The first one refers to the 3D dynamics of a bar,
represented by a Ferrers bar, while the second to a 3D spiral PERLAS potential
with
two arms. In both cases we have examined the phase space close to periodic
orbits of x1v1, which is a family introduced in the system as bifurcation of
the central family x1, at its vertical
2:1 resonance.
Orbits of this family are associated with the presence of a peanut, or X-shaped,
bulge in the side-on view of the Ferrers model \citep{psa02, pk14a} and with the
enhancement of the spiral arms in the PERLAS potential \citep{chvv19}.
Our main conclusions are the following:
\begin{enumerate}
 \item The structure of the phase space in the neighbourhood of successive
orbits of the x1v1 family in both models presents similar features, as the
stability of the family experiences a S$\rightarrow\!\!\Delta\!\!\rightarrow$S
transition with increasing \ej\!\!. The evolution of the phase space
structure can be summarized as follows:
\begin{itemize}
 \item Before the S$\rightarrow\!\!\Delta$ transition, the volume of regular
orbits around the stable representatives of x1v1 decreases with increasing
\ej\!\!. Approaching the
critical point, we have to decrease the perturbations we apply to one of the
four initial conditions in order to find in the Poincar\'{e} spaces of section
toroidal
surfaces with smooth colour variation on them. Simultaneously the
tori (as e.g. in Fig.~\ref{delta1_10pc}a) become flatter, tending to become
disky.
 \item Just beyond the S$\rightarrow\!\!\Delta$ transition, around the complex
unstable periodic orbits we find regular structures, namely disky confined
tori \citep{pf85a, pf85b, kpc11}. There is an internal structure on them, in the
sense that the consequents cover the surfaces of the confined tori following
specific spiral
patterns. The number of the arms of these spiral patterns varies.
  \item A next phase in the evolution of the phase space structure in
the
neighbourhood of the x1v1 periodic orbits, appears as
we depart from the S$\rightarrow\!\!\Delta$ transition point towards larger
energies, keeping the relative perturbation constant. We find then consequents
initially building spiral patterns with smooth colour variation along their
arms, which later diffuse in phase space building clouds of scattered points,
filling the available volume of the phase space, limitted by the surface of
zero velocity.
 \item For the largest part of a $\Delta$ region, integrating orbits in the
immediate neighbourhood of the x1v1 periodic orbits leads to clouds of
scattered points in the Poincar\'{e} cross sections. However, in many cases the
orbits remain confined during a significant period within a certain subset of
the 4-dimensional space. This plays a major role for practical
applications.
 \item Close to the $\Delta\!\!\rightarrow$S transition the phase space is
organized again, however within small volumes around the complex unstable
periodic orbit. Namely, we encounter again disky confined tori.
 \item Finally, beyond the $\Delta\!\!\rightarrow$S transition, in the region
where the family is again stable, the restoration of order has again a gradual
character. The radius within which we find regular orbits around the
periodic orbit increases with \ej\!\!.
\end{itemize}
\item The shrinking of the volume occupied by regular orbits around the
stable x1v1 periodic orbits and the evolution of the tori towards a disky
morphology as we approach the critical energy for a S$\rightarrow\!\!\Delta$
transition, has been encountered in all studied cases. This is the second case
we know that the deformation of a phase space structure close to a periodic
orbit as the energy varies, foretells an impending change of stability
\citep[the first case has been presented by][for changes in the topology of
invariant tori before a S$\rightarrow$U transition]{pk14a}.
\item Within an \ej interval in which the x1v1 family is
complex unstable, orbits in the neighbourhood of
periodic orbits with the same $\Delta$ (Eq.~\ref{delta}), subject to the same
amount of relative
perturbations, do not have the same degree of chaoticity. In the cases we
studied, they appear more chaotic in the ascending part of the U-type curve of
the (\ej\!\!,$\Delta$) diagrams, towards the critical point
of the $\Delta\rightarrow$S transition. In that respect, there is no perfect
symmetry in the phase space structures around complex unstable periodic
orbits
with the same $\Delta$.
\item We underline the role of the phase space environment around a
periodic
orbit for the determination of the behaviour of the perturbed orbits. In many
cases displacements of the initial conditions of a periodic orbit along a
certain direction, may
bring the initial conditions of the perturbed orbit in zones of influence of
other periodic orbits
(stable or unstable). The variation of the GALI$_2$ indices may warn us about
such cases.
\item In both models, many orbits eventually expressing a chaotic character are
structure-supporting within a 5~Gyr period. Especially for the spiral PERLAS
potential, we conclude that even in the larger complex unstable energy
interval, there are orbits relatively close to complex unstable periodic
orbits, which contribute
to the reinforcement of the spiral arms of the model for considerable time
intervals.
\item Supporting further the above conclusion, we note that the orbits
close to the periodic orbits of the small complex unstable energy intervals in
both models, can hardly be distinguished from regular during a 5~Gyr period.
\end{enumerate}

\vspace{0.5cm}
\noindent \textit{Acknowledgements}

We thank G.~Contopoulos and M.~Katsanikas for fruitful discussions and useful
comments. This work has been carried out in the frame of the project ``Numerical
investigation of the impact of Complex Instability to the phase space structure
of Dynamical Systems'' of the Research Center for Astronomy of the Academy of
Athens.
L.C.V thanks the Fondo Nacional de Financiamiento
para la Ciencia, La Tecnolog\'{\i}a y la innovaci\'{o}n "FRANCISCO JOS\'{E} DE
CALDAS",
MINCIENCIAS, and the VIIS for the economic support of this research. L.C.V
acknowledges the support of the postdoctoral Fellowship of DGAPA-UNAM, Mexico.
Ch.S.
acknowledges support by the Research Committee (URC) of the University of Cape
Town.



\printcredits

\bibliographystyle{cas-model2-names}

\bibliography{cas-refs}

%
%
%

\end{document}